\documentclass[a4paper,amsmath,amssymb,superscriptaddress,aps,prb,reprint,showpacs,nofootinbib]{revtex4-2}
\usepackage{xcolor,graphicx}
\usepackage{subfigure}
\usepackage{MnSymbol}
\usepackage{braket}
\usepackage{tabularx}
\usepackage{booktabs}
\usepackage{tikz}
\usepackage{pgfplots}
\usepackage[utf8]{inputenc}
\usepackage{floatrow}
\usepackage{comment}
\newfloatcommand{capbtabbox}{table}[][\FBwidth]

\begin{document}

\title{From linear to circular polarized light - Floquet engineering in Kitaev-Heisenberg materials with Lissajous figures}

\author{Pascal Strobel}

\affiliation{%
Institut f\"ur Funktionelle Materie und Quantentechnologien,
Universit\"at Stuttgart,
70550 Stuttgart,
Germany}

\author{Maria Daghofer}
\affiliation{%
Institut f\"ur Funktionelle Materie und Quantentechnologien,
Universit\"at Stuttgart,
70550 Stuttgart,
Germany}
\affiliation{Center for Integrated Quantum Science and Technology, University of Stuttgart,
Pfaffenwaldring 57, 
70550 Stuttgart, Germany}

\date{\today}

\begin{abstract}
This paper discusses Floquet engineering with arbitrary polarization in $\alpha$-RuCl$_3$. We describe the influence of arbitrary polarization and the limiting cases of linear and circular polarization. The corresponding model is derived via perturbation theory up to fourth-order. Starting from linear and circular polarization we bridge the gap between those two limiting cases. We then we study more complex Lissajous figures and general trends arising for them.
\end{abstract}

\maketitle
\section{Introduction}\label{intro}
Floquet engineering has proven itself to be a promising tool for tuning magnetic interactions \cite{RevModPhys.89.011004,Mentink_2017,Oka}. When applying light periodic in time the system can be described with a time independent effective Hamiltonian known as Floquet Hamiltonian~\cite{Mentink_2015,PhysRevLett.116.125301,Claassen2017,PhysRevLett.125.047201}. Light frequency and amplitude  then modify the system's intrinsic interactions. This makes modifying system properties via light feasible. It has been shown that the procedure of an approximate time-independent Hamiltonian is indeed reasonable for short but experimentally accessible timescales \cite{KUWAHARA201696}, before the system experiences heating and the Floquet description breaks down.

Mott insulators with strong spin-orbit coupling (SOC), which are assumed to be close to realizing a Kitaev model~\cite{Winter_2017}, seem to be a promising playground for Floquet engineering. In materials like $\alpha$-RuCl$_3$ a KSL was proposed, however the groundstate is likely an antiferromagnet with a zig-zag pattern~\cite{PhysRevB.92.235119}. Therefore the idea of altering the intrinsic interaction parameters arose. There have been a plethora of attempts~\cite{PhysRevB.96.205147,PhysRevB.97.245149,PhysRevB.103.L140402,PhysRevLett.118.187203,PhysRevLett.119.037201,PhysRevB.95.180411,PhysRevLett.119.227208,PhysRevLett.120.117204,PhysRevLett.123.237201,PhysRevMaterials.1.052001,PhysRevB.102.094407,PhysRevB.99.214410} to tune $\alpha$-RuCl$_3$ into the Kitaev spin liquid (KSL) phase, the most promising of which up to date is propably applying a magnetic field~\cite{PhysRevLett.118.187203,PhysRevLett.119.037201,PhysRevB.95.180411,PhysRevLett.119.227208,PhysRevLett.120.117204,zhou2022intermediate}.  

Most recently, there have been several proposals to tune Kitaev candidate materials into the sought after KSL via Floquet engineering. The focus here has either been on linearly polarized light~\cite{PhysRevB.104.214413,PhysRevB.105.085144}, where the light angle gives new possibilities to manipulate interactions, in addition to the light frequency, or circularly polarized light which has a isotropic influence on the system~\cite{PhysRevB.103.L100408}. For circular polarization, an induced inverse Farraday effect~\cite{banerjee2021inverse} has been discussed, which is found when ligands are explicitly included via third-order perturbation theory. This suggests that including ligands leads to novel tuning possibilities~\cite{Kumar2022,PhysRevResearch.4.L032036}. Up until now the transition from circularly to linearly polarized light has been lacking in the context of Kitaev-Heisenberg materials. An effective model to describe arbitrary polarizations has been introduced in~\cite{PhysRevB.94.235419,PhysRevLett.126.177201,PhysRevB.103.134435}. Starting from there, there have been proposals of tuning e.g. Cd$_3$As$_2$ via Lissajous figures~\cite{PhysRevLett.128.066602}.

In this paper we want to transfer the methods of~\cite{PhysRevLett.126.177201,PhysRevB.103.134435,PhysRevLett.128.066602} to candidate Kitaev-Heisenberg models, specifically to the most promising $\alpha$-RuCl$_3$. Our goal is to explore new tuning possibilities via Lissajous figures as well as bridge the gap between linearly polarized (LPL) and circularly polarized light (CPL). 

To do so we derive an expression of the Peierls substitution for arbitrary polarizations. From there we obtain an effective model via perturbation theory up to fourth-order in order to capture the formerly mentioned inverse Farraday effect and also yet lacking contributions from fourth-order terms. We include fourth-order terms, since Kitaev interactions already have fourth-order contributions in the absence of light. Projecting this effective Hamiltonian into the $j=1/2$-basis then yields an effective Floquet-Kitaev-Heisenberg model capturing the influence of arbitrary polarized light. This model reproduces the results of the LPL~\cite{PhysRevB.105.085144,PhysRevB.104.214413} and CPL Hamiltonians~\cite{PhysRevResearch.4.L032036,Kumar2022} in the respective limit. We find that that third and fourth-order perturbation theory not only gives rise to the additional magnetic field for CPL, but also yields additional interactions for LPL. With this model we study the already known limiting cases and proceed by investigating more complex Lissajous figures in order to find new pathways for tuning Kitaev-Heisenberg materials. In this process we try to carve out the role of the multiple variables like amplitude, frequency, etc. arising in the model.
    
This paper is structured as follows:
In Sec.\ref{theo} we derive the effective model for arbitrary polarized light. We first expand the model of~\cite{PhysRevB.105.085144} in order to obtain the second-order terms. 
And then calculate third and fourth-order terms for arbitrary polarizations, similar to~\cite{Kumar2022}. We then give an expression for all interactions arising, and briefly discuss the nature of these. 
In Sec.\ref{LC} we analyze the model derived in Sec.\ref{theo}. Focus is first on light with a phase shift of $\epsilon=0$ and $\epsilon=\pi/2$ [see (EQ)] in Sec.\ref{LCA}. Here we study a frequency multiplicity of $N=1$ [see (EQ)], representing CPL and LPL at a driving frequency significantly above any resonances. We compare our results to the results of~\cite{Kumar2022} in order to confirm the validity of our model. 
In Sec.\ref{LCB} we treat the phase shift $\epsilon$ as a free parameter. We then continuously go from CPL to LPL, and study the change of the interaction parameters. Building upon the results of Sec.\ref{LCA} we start with the same driving frequency. In order to assess the influence of driving frequency we also study the case of driving in between resonances of the effective Floquet Hamiltonian.
Last but not least in Sec.\ref{LCC} we investigate the influence of the second Lissajous parameter, the frequency multiplicity $N$. 
We conclude and summarize the results in Sec.\ref{sum}.  

\section{Theory and Methods}\label{theo}
\subsection{Floquet Hamiltonian}\label{theoA}
As shown in~\cite{PhysRevB.105.085144} one can describe the kinetic part of the Hamiltonian with a Hubbard Hamiltonian where the hopping part is modulated via the Peierls substitution $t\rightarrow t\mathcal{B}(\vartheta,\mathbf{A}(\tau))$. The Hamiltonian for the $z$-bonds can then be written as
\begin{align}
H^z_{\mathrm{kin}}(\tau)=&-\sum_{\sigma,\braket{ij}_{z}}\mathcal{B}(0,\mathbf{A})\mathbf{d}_{i,\sigma}^{\dagger}\mathbf{T}^{z}\mathbf{d}^{\phantom{\dagger}}_{j,\sigma}\notag\\&+t_{pd}\,\bigg[\mathcal{B}\left(-\frac{\pi}{4},\mathbf{A}\right)\left(d_{zx,i,\sigma}^{\dagger}p_{1,\sigma}^{\phantom{\dagger}}-p_{2,\sigma}^{\dagger}d_{yz,j,\sigma}^{\phantom{\dagger}}\right)\notag\\
&+\,\mathcal{B}\left(\frac{\pi}{4},\mathbf{A}\right)\left(d_{zx,i,\sigma}^{\dagger}p_{2,\sigma}^{\phantom{\dagger}}-p_{1,\sigma}^{\dagger}d_{yz,j,\sigma}^{\phantom{\dagger}}\right)+\mathrm{h.c.}\bigg],
\label{eq:EQ1}
\end{align}
where $\mathbf{d}_{i,\sigma}$ ($\mathbf{d}_{i,\sigma}^{\dagger}$) denotes the vector of annihilation (creation) operators $d_{i\alpha\sigma}$ ($d_{i\alpha\sigma}^{\dagger}$) that annihilate (create) an electron in the $d$-orbital  $\alpha\in [xy,yz,zx]$ on site $i$ with spin $\sigma$, time $\tau$. The matrix $\mathbf{T}^z$ contains direct $d$-$d$ hopping modified by the Peierls substitution $\mathcal{B}(0,\mathbf{A})$. $p_{i,\sigma}$ ($p_{i,\sigma}^{\dagger}$) annihilates (creates) a $p$-electron on the ligand ion, and $t_{pd}$ is the hopping between $d$- and $p$-states. The kinetic Hamiltonian in $y$- and $x$-direction can be derived by adjusting $\vartheta$ and choosing the corresponding hopping processes $\mathbf{T}^{\gamma}$~\cite{PhysRevB.93.214431,PhysRevB.105.085144}  as well  $t_{pd}$.
We explicitly included the hopping processes from the $d$-orbitals to the $p$-ligand since it has been shown~\cite{Kumar2022,PhysRevResearch.4.L032036} that they can not be simply be integrated out. 
Since we want to consider arbitrary polarization we describe the vector-potential as 
\begin{align}
\mathbf{A}=\begin{pmatrix}\frac{E_x}{\omega}\sin(\omega t)\\ \frac{E_y}{N\omega}\sin(N\omega t+\epsilon)\end{pmatrix}\label{eq:EQ1.5}
\end{align}
 giving rise to a Peierls substitution of the form
\begin{align}
\mathcal{B}(\vartheta,\mathbf{A})=&\mathrm{exp}\bigg[i\mathbf{A}
\begin{pmatrix}
\cos(\vartheta)\\
\sin(\vartheta)
\end{pmatrix}
\bigg],
\label{eq:EQ2}
 \end{align}
which captures all possible Lissajous figures via the tuning parameters $N$ and $\epsilon$. $N$ is the ratio between the frequency in $x$-direction $\omega$ and the frequency in $y$-direction $N\omega$ and epsilon describes the phase shift between $x$- and $y$-polarized light. Fixing $N=1$, $\epsilon$ gradually shifts polarization from circular ($\epsilon=\pi/2$) to linear ($\epsilon=0$). Choosing $N>1$ yields Lissajous figures that have not been investigated in the context of $\alpha$-RuCl$_3$ so far. Unlike in \cite{PhysRevB.105.085144} focus on $\varphi=\pi/4$ as angle of LPL, i.e. $E_x=E_y=E_0$.

The on-site interactions are captured with the Kanamori Hamiltonian
\begin{align}
H_{\mathrm{int}}=&U\sum_{i,\alpha}n_{i\alpha\uparrow}n_{i\alpha\downarrow}+U'\sum_{i,\sigma}\sum_{\alpha<\beta}n_{i\alpha\sigma}n_{i\beta \,-\sigma}\notag\\   
   &-J_H\sum_{i,\alpha\neq\beta}(d_{i\alpha\uparrow}^{\dagger}d_{i\alpha\downarrow}d_{i\beta\downarrow}^{\dagger}d_{i\beta\uparrow}-d_{i\alpha\uparrow}^{\dagger}d_{i\alpha\downarrow}^{\dagger}d_{i\beta\downarrow}d_{i\beta\uparrow})\notag\\
   &+(U'-J_H)\sum_{i,\sigma}\sum_{\alpha<\beta}n_{i\alpha\sigma}n_{i\beta\sigma}+\Delta\sum_{i,\sigma}p^{\dagger}_{i\sigma}p^{\phantom{\dagger}}_{i\sigma},
   \label{eq:EQ3}
\end{align}
with intraorbital interaction $U$, interorbital interaction
$U'=U-2J_H$, Hund’s coupling $J_H$, and density $n_{i\alpha\sigma}$.

The complete Hamiltonian is periodic in time. It has been shown with Floquet's theorem~\cite{PhysRev.138.B979}, that such Hamiltonians can be described with a time independent Floquet Hamiltonian~\cite{PhysRevLett.126.177201}. In our case, the Hamiltonian takes the form
\begin{align}
H^{F}&=-\sum_{l,n}H_{\mathrm{kin}}^{l}\ket{n+l}\bra{n}+\sum_{l}\left(H_{\mathrm{int}}+l\omega\right)\ket{l}\bra{l},
 \label{eq:EQ4}
\end{align}
where $l,m$ is the number of photons. $H_{\mathrm{kin}}^{l}$ describes a hopping process with the absorption of $l$ photons and can be derived via a Brillouin-Wigner expansion, as in~\cite{PhysRevB.94.235419},
\begin{align}
H_{\mathrm{kin}}^{l}=\frac{1}{2\pi}\int_0^{2\pi}H_{\mathrm{kin}}(\tau)e^{il\omega\tau}\text{d}\tau.
 \label{eq:EQ5}
\end{align}
Looking at (\ref{eq:EQ1}), we see that the Peierls substitution $\mathcal{B}$ is the only time dependent term in $H_{\mathrm{kin}}(\tau)$. Performing the integration for this term yields
\begin{align}
\mathcal{B}_{l}(\vartheta,\mathbf{A})&=\frac{1}{2\pi}\int_0^{2\pi}\mathcal{B}_N(\vartheta,\mathbf{A})e^{il\omega\tau}d\tau\notag\\
&=\sum_n\mathcal{J}_{l-Nn}\left(\frac{E_0}{\omega}\cos(\vartheta)\right)\mathcal{J}_{n}\left(\frac{E_0}{N\omega}\sin(\vartheta)\right)e^{i\epsilon n},
\label{eq:EQ6}
\end{align}
where we used the Jacobi-Anger expansion~\cite{PhysRevLett.121.107201} before integration. The Floquet-Hamiltonian therefore has the exact same hopping processes as the bare Hamiltonian, but with dressed hopping strengths that now depend on $\epsilon,N, A_0,$ and $\alpha$.

\subsection{Second-order perturbation theory}\label{theoB}
$\alpha$-$\mathrm{RuCl}_3$ is considered to be  a Mott insulator, where the Coulomb repulsion $U$ is much larger than the hopping parameters $\mathbf{T}^{\gamma}$ and $t_{pd}$ of $H_{\mathrm{kin}}^{F}$. Therefore we can treat the kinetic part of the Floquet Hamiltonian (\ref{eq:EQ4}) in perturbation theory. The most conventional approach is to calculate an effective second-order Hamiltonian
\begin{align}
H^{F}_{\mathrm{eff}}&=\sum_{l,\alpha}\frac{H_{\mathrm{kin}}^{-l}\ket{\Psi_{\alpha}^{d}}\bra{\Psi_{\alpha}^{d}}H_{\mathrm{kin}}^{l}}{E_{\alpha}+l\omega},
\label{eq:EQ7}
\end{align}
with $l$ the number of absorbed(emitted) photons, $\omega$ the driving frequency and $\ket{\Psi_{\alpha}^{d}}$ the manifold of states with a double occupation on one site.
Projecting the effective Kugel-Khomskii type Floquet Hamiltonian into the $j=1/2$ basis then gives rise to a Kitaev-Heisenberg model like in~\cite{PhysRevLett.112.077204}. 

The distinct feature of the Floquet-Kitaev-Heisenberg Hamiltonian is the dependence of the interaction parameters on $\omega$ and $E_0$ for CPL, and additionally on $\varphi$ for LPL. In order to obtain the second-order Hamiltonian for arbitrary polarizations one has to simply exchange the Bessel functions in~\cite{PhysRevB.105.085144} with expression derived in (\ref{eq:EQ6}). The interaction parameters obtained are given in App.~\ref{AA}. 

In these calculations the hopping over the $p$-ligand atoms is integrated out and is included in the $t_2$ hopping $t_2\rightarrow t_2+t_{pd}/\Delta$. While this is valid in case of systems without driving~\cite{PhysRevLett.112.077204}, a light field induces a complex phase for each hopping, which precludes including $t_{pd}$ into the $t_2$ hopping. We therefore have to calculate virtual $t_{pd}$-hopping strengths explicitly, necessitating perturbation theory up to fourth-order.  
\subsection{Third-order Perturbation theory}\label{theoC}
Third-order perturbation theory considers hopping processes, where one $d\rightarrow d$ process is mediated by a ligand $p$-atom, i.e. occors along a $d$-$p$-$d$ path. For CPL there have been several theoretical articles proposing an inverse Faraday effect arising due to these additional hopping processes~\cite{Kumar2022,PhysRevResearch.4.L032036} which is also supported by recent experimental results~\cite{PhysRevResearch.4.L032032}. Furthermore~\cite{Kumar2022} introduced analytic expressions for the Kitaev-Heisenberg Hamiltonian in third-order perturbation theory for CPL.

In this chapter we build on these findings and extend them to arbitrary polarization, deriving analytic expressions for third-order correction terms. The third-order contributions to the effective Floquet Hamiltonian can be calculated via 
\begin{align}
H^{F}_{\mathrm{eff}}&=\sum_{l,m}\sum_{\beta,\alpha}\frac{H_{\mathrm{kin}}^{-l-m}\ket{\Psi_{\alpha}^{d}}\bra{\Psi_{\alpha}^{d}}H_{\mathrm{kin}}^{m}\ket{\Psi_{\beta}^{p}}\bra{\Psi_{\beta}^{p}}H_{\mathrm{kin}}^{l}}{\left(E_{\alpha}+(m+l)\omega\right)\left(\Delta+l\omega\right)}, \label{eq:EQ8}
\end{align}
where $\ket{\Psi_{\alpha}^{d}}(\ket{\Psi_{\alpha}^{p}})$ is the excited-state manifold with an additional electron in a $d$-($p$-)orbital with excitation energies $E_{\alpha}$($\Delta$). The possible energies for two electrons in a $t_{2g}$ $d$-shell are $E_P=U-3J_H$, $E_D=U-J_H$, and $E_S=U+2J_H$. In contrast to second-order processes, where $l$ photons get absorbed(emitted) in the first hopping process and emitted (absorbed) in the second hopping process, in the third-order process $l$ photons get absorbed (emitted) in the $d\rightarrow p$ hopping process and another $m$ photons get absorbed (emitted) in the $p\rightarrow d$ hopping which then get collectively emitted (absorbed) in the $d\rightarrow d$ process so that no photons are present in the final state. 

Using the projections of~\cite{kumar2021floquet} our calculations for arbitrary polarization (see more details in App.~\ref{AC}) give rise to three new interaction terms in addition to the $J$, $K$ and $\Gamma$ terms present in systems without driving. The first one is the already reported magnetic field term $h^3$ arising from broken time reversal symmetry and inducing the inverse Faraday effect. The other two terms break inversion symmetry (IS) ($D$) and induce further anisotropies ($\mu$). The full third-order Hamiltonian projected onto the $j=1/2$ basis then reads    
\begin{align}
H^3_{\mathrm{eff}}=&\sum_{\gamma,\braket{i,j}_{\gamma}}J^3_{\gamma}\mathbf{S}_i\mathbf{S}_j+K^3_{\gamma}S^{\gamma}_iS_j^{\gamma}+\Gamma^3_{\gamma}\left(S_i^{\alpha}S_j^{\beta}+S_i^{\beta}S_j^{\alpha}\right)\notag\\
&+D^3_{\gamma}\mathbf{e}_{\gamma}\left(\mathbf{S}_i\times\mathbf{S}_j\right)+\mu^3_{\gamma}\left(S_i^{\alpha}S_j^{\alpha}-S_i^{\beta}S_j^{\beta}\right)+h^3_{\gamma} \left( S_i^{\gamma}+S_j^{\gamma}\right)
\label{eq:EQ9}
\end{align}
the terms for the $z$-bond are given as 
\begin{widetext}
\begin{align}
K^3_z=&\sum_{m,l}\frac{t_{pd}^2}{\Delta+m\omega}\bigg[ \mathrm{Re}\left(\mathfrak{B}^3_{l,m}+\mathfrak{B}^3_{m,l}\right)\frac{12}{9}\left(\frac{t_2}{E_D+(l+m)\omega}-\frac{t_2}{E_P+(l+m)\omega}\right)\notag\\
&+\mathrm{Im}\left(\mathfrak{B}^3_{l,m}-\mathfrak{B}^3_{m,l}\right)\frac{8}{27}\left(\frac{t_1-t_3}{E_D+(l+m)\omega}+\frac{2t_1+t_3}{E_S+(l+m)\omega}+\frac{6t_2}{E_P+(l+m)\omega}\right)\bigg]\label{eq:EQ10}\\
\Gamma^3=&\sum_{m,l}\frac{t_{pd}^2}{\Delta+m\omega}\mathrm{Re}\left(\mathfrak{B}^3_{l,m}+\mathfrak{B}^3_{l,m}\right)\frac{4}{9}\left(\frac{t_1-t_3}{E_P+(l+m)\omega}-\frac{t_1-t_3}{E_D+(l+m)\omega}\right)\label{eq:EQ11}\\
\mu^3_z=&\sum_{m,l}\frac{-t_{pd}^2}{\Delta+m\omega}\mathrm{Re}\left(\mathfrak{B}^3_{l,m}-\mathfrak{B}^3_{m,l}\right)\frac{4}{9}\left(\frac{t_2}{E_P+(l+m)\omega}+\frac{t_2}{E_D+(l+m)\omega}\right)\label{eq:EQ12}\\
D^3_z=&\sum_{m,l}\frac{t_{pd}^2}{\Delta+m\omega}\mathrm{Re}\left(\mathfrak{B}^3_{l,m}-\mathfrak{B}^3_{m,l}\right)\frac{8}{27}\left(\frac{2t_1+t_3}{E_S+(l+m)\omega}+\frac{t_1-t_3}{E_D+(l+m)\omega}+\frac{3(t_1+t_3)}{E_P+(l+m)\omega}\right)\label{eq:EQ13}\\
h^3_z=&\sum_{m,l}\frac{-t_{pd}^2}{\Delta+m\omega}\mathrm{Im}\left(\mathfrak{B}^3_{l,m}-\mathfrak{B}^3_{m,l}\right)\frac{2}{9}\left(\frac{t_1-t_3}{E_D+(l+m)\omega}+\frac{t_1-t_3}{E_P+(l+m)\omega}\right),
\label{eq:EQ14}
\end{align}
\end{widetext}
with
\begin{align}
\mathfrak{B}^3_{l,m}=&\mathcal{B}_{-l-m}\left(0,\mathbf{A}\right)\mathcal{B}^{\ast}_{-l}\left(\frac{\pi}{4},\mathbf{\tilde{A}}\right)\mathcal{B}^{\ast}_{-m}\left(-\frac{\pi}{4},\mathbf{\tilde{A}}\right),
\label{eq:EQ15}
\end{align}   
where $\mathcal{B}_{l}$ is introduced in (\ref{eq:EQ6}) and $\mathbf{\tilde{A}}=\mathbf{A}/\sqrt{2}$. It is of note that in third-order perturbation theory there are no contributions to the Heisenberg term $J$, contrary to the findings of~\cite{Kumar2022}.\footnote{We believe the reason for that is the negligence of $d\rightarrow d\rightarrow p\rightarrow d$ hopping processes in~\cite{Kumar2022} which lead to the vanishing of $J^3$ and a finite $D^3$.} Results for the $x$- and $y$-bond can be deduced from (\ref{eq:EQ10})-(\ref{eq:EQ14}) by selecting $\vartheta$ in (\ref{eq:EQ15}) accordingly.
\subsection{Fourth-order perturbation theory}\label{theoD}
\begin{figure}
\includegraphics[width=\columnwidth]{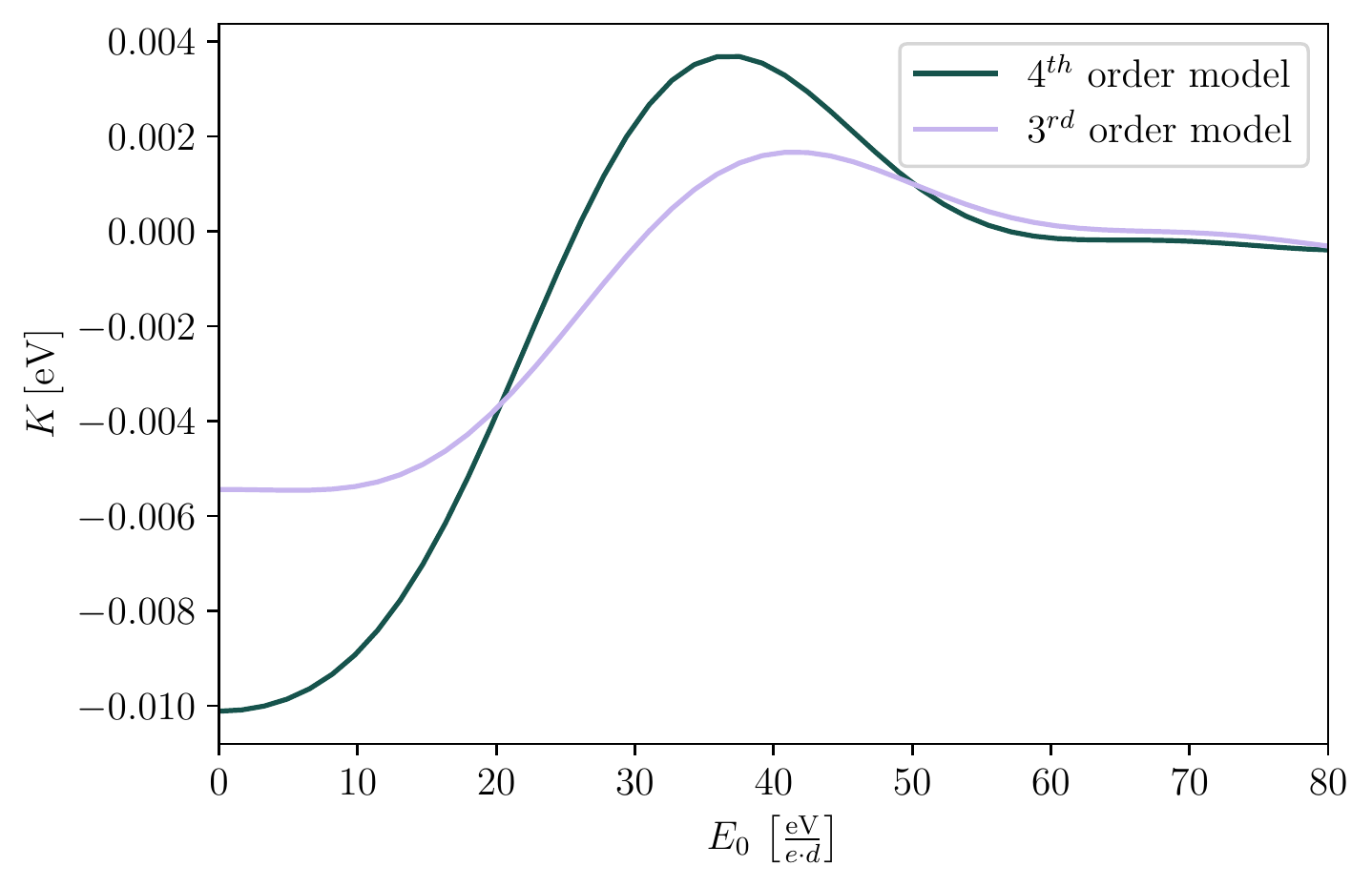}
\caption{Comparison of the Kitaev interaction in dependency of $E_0$ at $\omega=12.0\,$eV obtained with the third order model (mauve) and the fourth order model (dark green).}
\label{fig:Fig1}
\end{figure}
While the inverse Faraday might be captured completely via third-order perturbation theory, the Kitaev interaction still lacks significant contributions that come from fourth-order contributions. As is evident from the conventional Kitaev-Heisenberg model~\cite{PhysRevLett.112.077204}  $t_{pd}^4$ terms, i.e. $d$-$p$-$d$-$p$-$d$ processes, are the driving force for a sizable Kitaev term. It is therefore crucial to include terms where both $d\rightarrow d$ hopping processes are mediated by ligand $p$-atoms explicitly. This can be done in a similar manner as for second (\ref{eq:EQ7}) and third-order terms (\ref{eq:EQ8}). The contributions to the interaction terms from fourth-order perturbation theory are then given by  
\begin{widetext}
  \begin{align}
  J^4=&\sum_{n,l,m,k}\frac{t_{pd}^4\,\delta_{n,-l-k-m}\left(\mathfrak{B}^{4}_{n,l}-\mathfrak{B}^{4}_{l,n}\right)\left(\mathfrak{B}^{4\,\ast}_{-m,-k}-\mathfrak{B}^{4\,\ast}_{-k,-m}\right)}{(\Delta+(l+m+k)\omega)(\Delta+m\omega)}\frac{2}{27}\left(\frac{1}{E_D+(m+l)\omega}+\frac{3}{E_P+(l+m)\omega}+\frac{2}{E_S+(l+m)\omega}\right)\label{eq:EQ16}\\
  K^4=&\sum_{n,l,m,k}\frac{t_{pd}^4\,\delta_{n,-l-k-m}}{(\Delta+(l+m+k)\omega)(\Delta+m\omega)}\bigg[\frac{2}{3}\left(\frac{1}{E_P+(l+m)\omega}-\frac{1}{E_D+(l+m)\omega}\right)\left(\mathfrak{B}^{4}_{l,n}\mathfrak{B}^{4\,\ast}_{-m,-k}+\mathfrak{B}^{4}_{n,l}\mathfrak{B}^{4\,\ast}_{-k,-m}\right)\notag\\
  &-\frac{2}{27}\left(\frac{2}{E_S+(l+m)\omega}+\frac{3}{E_P+(l+m)\omega}+\frac{4}{E_D+(l+m)\omega}\right)\left(\mathfrak{B}^{4}_{n,l}-\mathfrak{B}^{4}_{l,n}\right)\left(\mathfrak{B}^{4\,\ast}_{-m,-k}-\mathfrak{B}^{4\,\ast}_{-k,-m}\right)\bigg]\label{eq:EQ17}\\
  \mu^4=&\sum_{n,l,m,k}\frac{t_{pd}^4\,\delta_{n,-l-k-m}\left(\mathfrak{B}^{4}_{n,l}\mathfrak{B}^{4\,\ast}_{-m,-k}-\mathfrak{B}^{4}_{l,n}\mathfrak{B}^{4\,\ast}_{-k,-m}\right)}{(\Delta+(l+m+k)\omega)(\Delta+m\omega)}\frac{2}{18}\left(\frac{1}{E_D+(l+m)\omega}-\frac{1}{E_P+(l+m)\omega}\right)\label{eq:EQ18}  
  \end{align}
  \end{widetext}
  where $l,m,k$ are the photons absorbed and emitted in the virtual hopping process and the fourth-order equivalent of (\ref{eq:EQ15}) is given as  
\begin{align}
\mathfrak{B}^4_{n,l}=&\mathcal{B}_n\left(\frac{\pi}{4},\mathbf{\tilde{A}}\right)\mathcal{B}_l\left(-\frac{\pi}{4},\mathbf{\tilde{A}}\right).
\label{eq:EQ19}
\end{align}  
As expected for forth order we have non zero Kitaev interactions. In addition, there are contributions to Heisenberg and $\mu$ interactions. The absence of $h^4$ terms explains the remarkably good agreement of the third-order $h$ term in~\cite{Kumar2022} with the numerical results. 

Summarizing for arbitrary polarization there arise two new type of interactions in addition to the formerly known J,G, $\Gamma$ and h terms, which both break the IS of the system ($D$) and induce further anisotropies ($\mu$). In addition to that we found that Heisenberg interactions do not have a third-order contribution but a fourth-order contribution. This in tandem with Kitaev interactions having fourth-order contributions makes the inclusion of fourth-order terms a necessity. To showcase the influence of fourth order terms in $\alpha$-RuCl$_3$ we calculated Kitaev interactions in dependency of the driving amplitude $E_0$ both in third and fourth order. We use the same parameters as~\cite{Kumar2022}, i.e., \textit{ab initio}~\cite{PhysRevB.93.155143} and photoemission~\cite{Sinn2016}, in order to compare our results with~\cite{Kumar2022}. These parameters are used for the remainder of this paper. The results are displayed in Fig.\;\ref{fig:Fig1}. We notice, as already discussed, a significant difference at $E_0=0$, with Kitaev interactions in fourth order being significantly stronger than third order results. For finite $E_0$ the qualitative behavior is very comparable for $3^{rd}$- and $4^{th}$-order calculations, with a maximum at $E_0\approx 40\,\mathrm{eV}/(e\cdot d)$ and a strong suppression of Kitaev interactions for $E_0> 50\,\mathrm{eV}/(e\cdot d)$. However there is still a sizable difference in magnitude for finite $E_0$ between $3^{rd}$- and $4^{th}$-order results throughout the parameter range considered. This in combination with the difference at $E_0=0$ is clear evidence for the importance of $4^{th}$-order terms.

\section{Results - From linear to circular polarized light}\label{LC}

\subsection{Limiting cases $\epsilon=0$ and $\epsilon=\pi/2$}\label{LCA}
\begin{figure}
\subfigure[LPL $\epsilon=0$]{\label{fig:Fig2a}\includegraphics[width=\columnwidth]{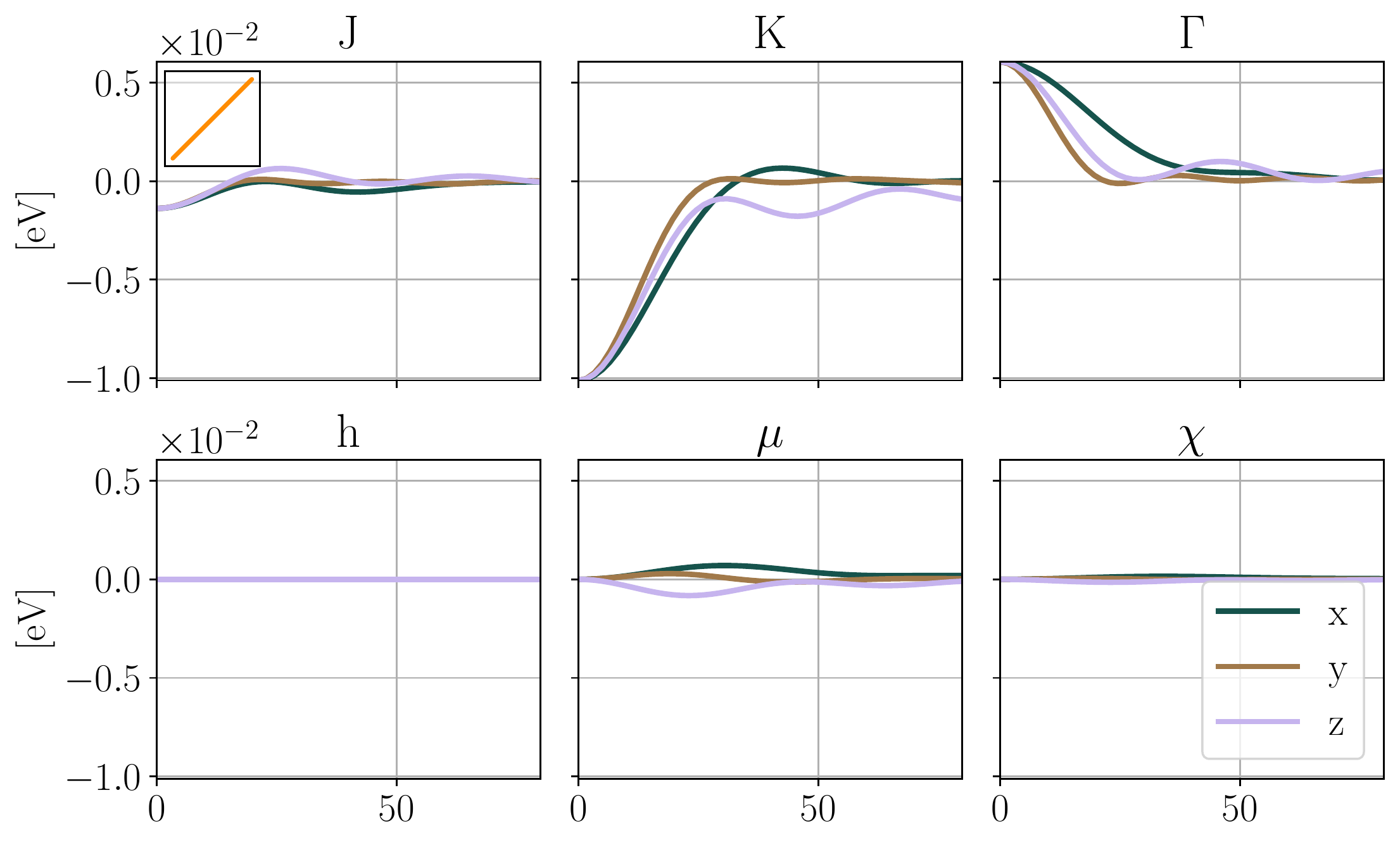}}
 \subfigure[CPL $\epsilon=\pi/2$]{\label{fig:Fig2b}\includegraphics[width=\columnwidth]{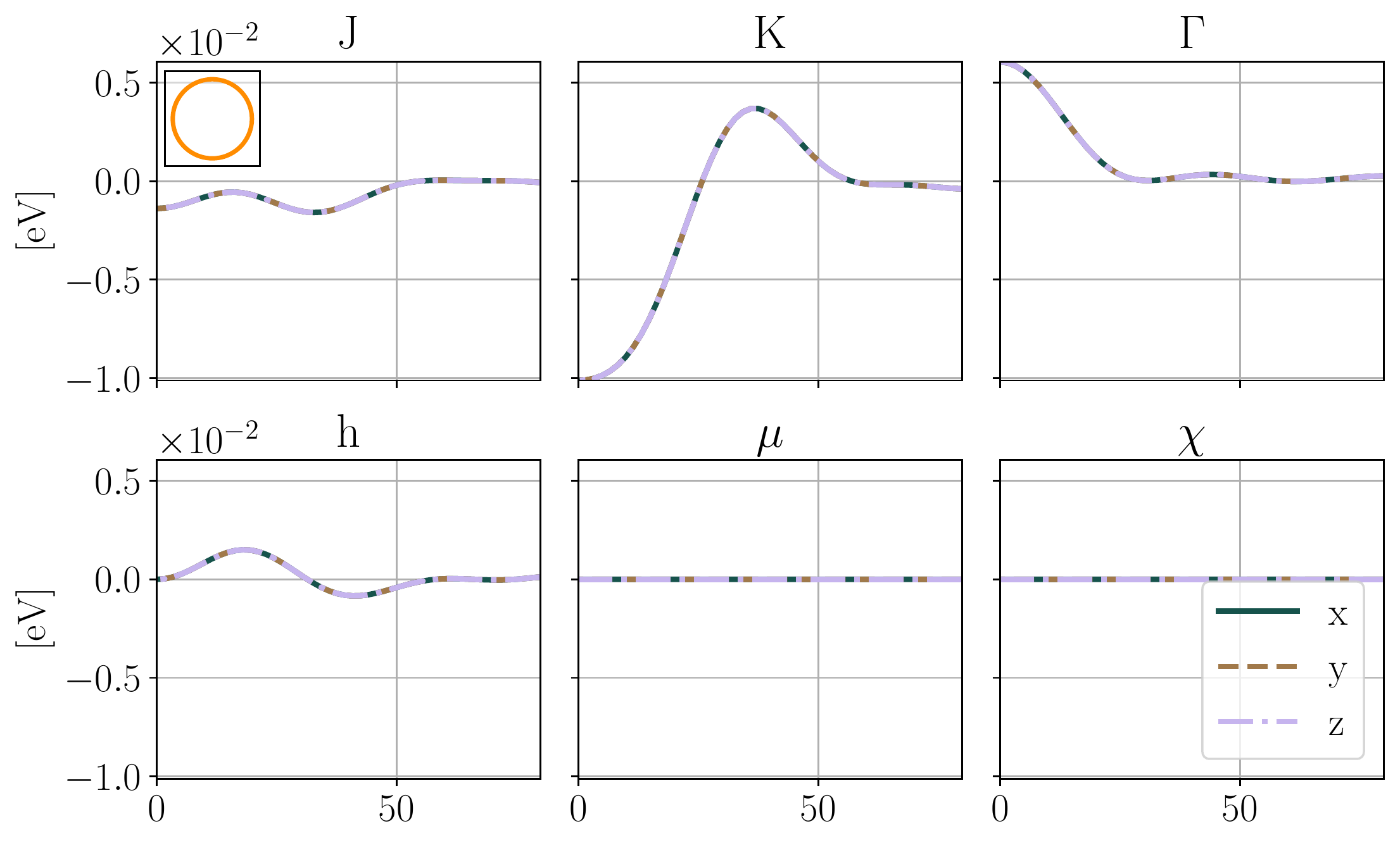}}
\caption{J-, K-, $\Gamma$-, $D$-, $\mu$-, and h-interactions in $x$-, $y$-, and $z$-direction in dependency of light amplitude $E_0$ at $\omega=12.0\,$eV. Displayed are the results for frequency multiplicity N=1}\label{fig:Fig2}
\end{figure}

We start our analysis of Lissajous figures with CPL and LPL, i.e. $\epsilon =\pi /2$ and $\epsilon =0$. The results for $\omega=12.0\,$eV are displayed in Fig.\ref{fig:Fig2a} and \ref{fig:Fig2b} for LPL and CPL respectively. Looking at Fig.\ref{fig:Fig2a} we immediately note the anisotropic influence of light on the different bond directions for all interactions as already reported in~\cite{PhysRevB.105.085144}.  In addition, we observe no induced magnetic field $h$ for any considered amplitudes $E_0$. Contrary to CPL, LPL does not break time reversal symmetry (TRS) and therefore does not induce a magnetic field. However the terms $D$ and $\mu$ have finite contributions for a nonzero amplitude $E_0$ and therefore IS is broken. These terms have not been reported in~\cite{PhysRevB.105.085144}, because they arise only in third and higher orders as discussed in Sec.\ref{theoC}. We therefore find that including third and higher orders explicitly is essential for arbitrary polarization. 

The results for CPL are displayed in Fig.\ref{fig:Fig2b}. Since CPL affects all bond directions in the same manner $x$-,$y$-, and $z$-interactions coincide. As already reported in~\cite{PhysRevResearch.4.L032036} there is a finite induced magnetic field pointing in the $\mathbf{n}=(1,1,1)$ direction. Meanwhile $D$ and $\mu$ vanish for CPL light, which was already evident from (\ref{eq:EQ12}), (\ref{eq:EQ13}), and (\ref{eq:EQ18}). We want to report that the results from our analytical expressions (\ref{eq:EQ10})-(\ref{eq:EQ14}) and (\ref{eq:EQ16})-(\ref{eq:EQ18}), with fourth-order terms and third-order terms appear to fit the ED results from~\cite{Kumar2022} far better than the analytical results of~\cite{Kumar2022}.\footnote{We attribute the small differences for $\omega=12.0\,$eV to a mistake in the indeces in (23) in the supplemental material of~\cite{Kumar2022}}. Especially the Kitaev term with the sizable positive values around $E_0\approx 40\,\mathrm{eV}/(e\cdot d)$ is remarkably close to the numerical results. Even our third order results have a good qualitative agreement with the numerical results as can be seen in Fig.\ref{fig:Fig1}.

Summarizing for the two limiting cases of CPL and LPL we find that linear polarizations induces the terms $D$ and $\mu$, which to our knowledge have not been reported previously. In addition LPL causes a bond anisotropy in dependence of the light angle, which is discussed in more detail in~\cite{PhysRevB.105.085144}. CPL breaks TRS and therefore induces a magnetic field, while keeping bond interactions isotropic.

\subsection{Elliptical Lissajou Figures (N=1)}\label{LCB}
\begin{figure}
\subfigure[J-, K-, and $\Gamma$-interactions]{\label{fig:Fig3a}\includegraphics[width=\columnwidth]{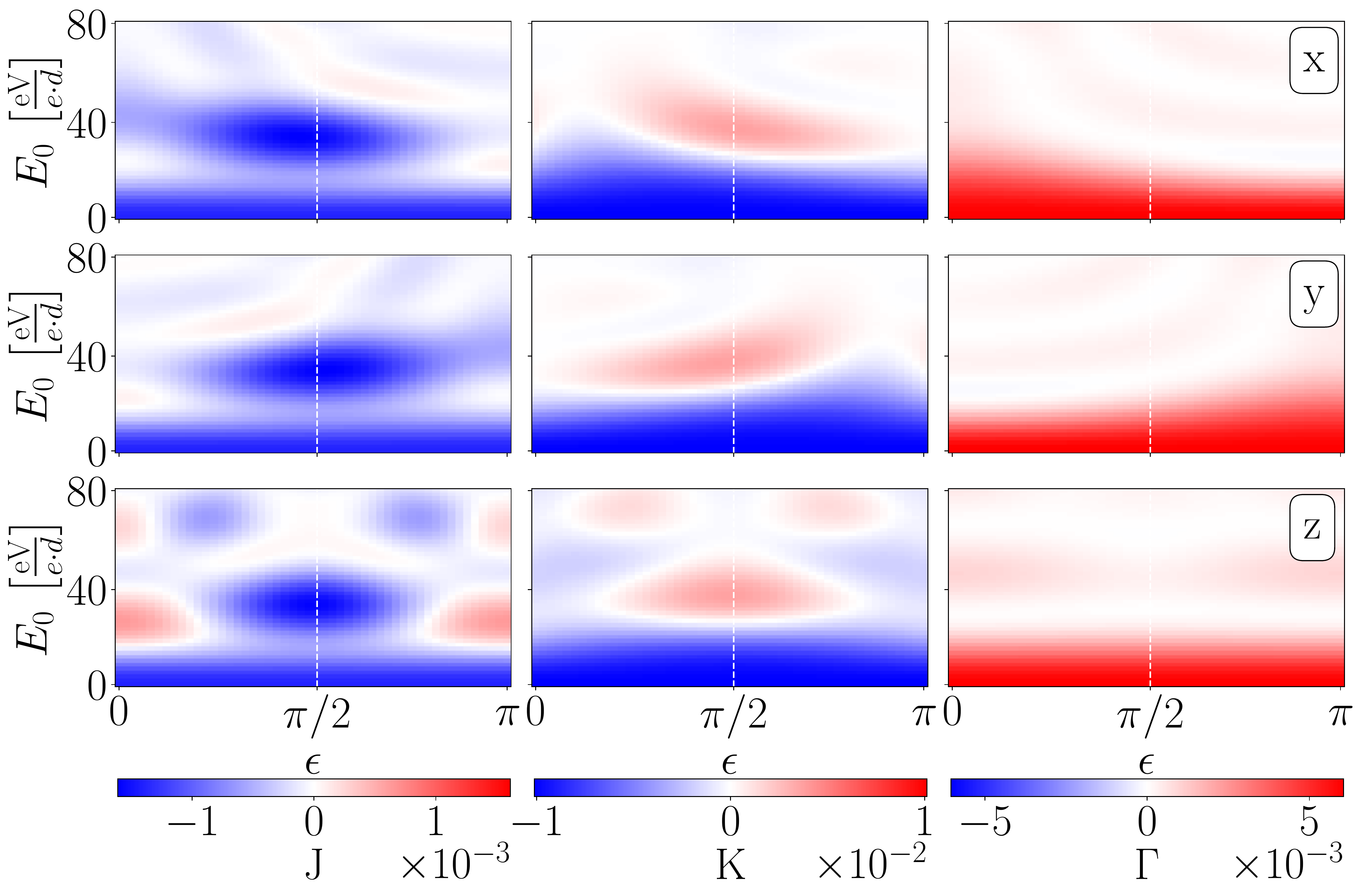}}
\subfigure[$D$-, $\mu$-, and h-interactions]{\label{fig:Fig3b}\includegraphics[width=\columnwidth]{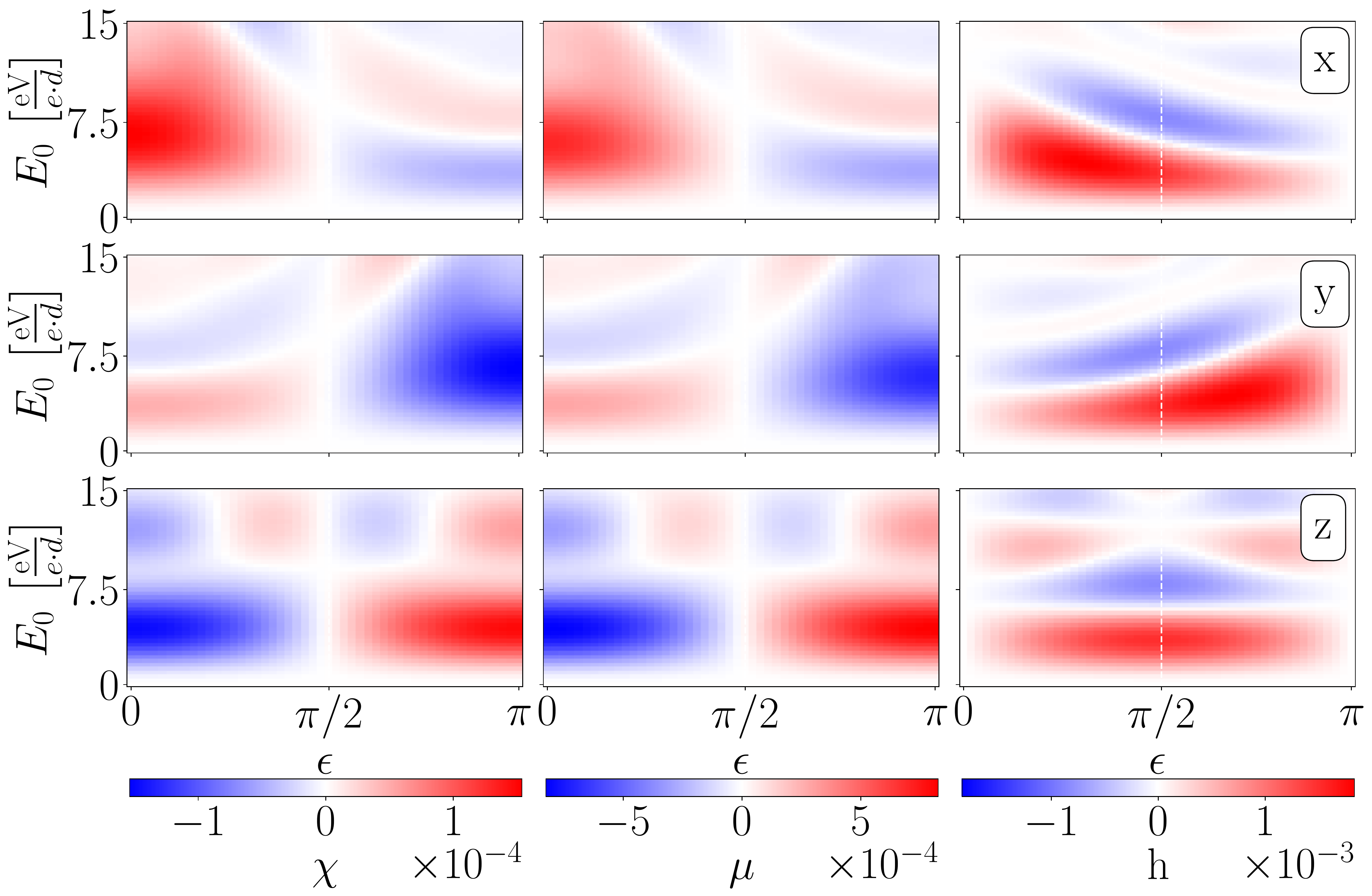}}
\caption{Interactions  in $x$-, $y$-, and $z$-direction in dependency of light amplitude $E_0$ and phase shift $\epsilon$ at $\omega=12.0\,$eV. Displayed are the results for frequency multiplicity N=1}\label{fig:Fig3}
\end{figure} 
\begin{figure}
\subfigure[J-, K-, and $\Gamma$-interactions]{\label{fig:Fig4a}\includegraphics[width=\columnwidth]{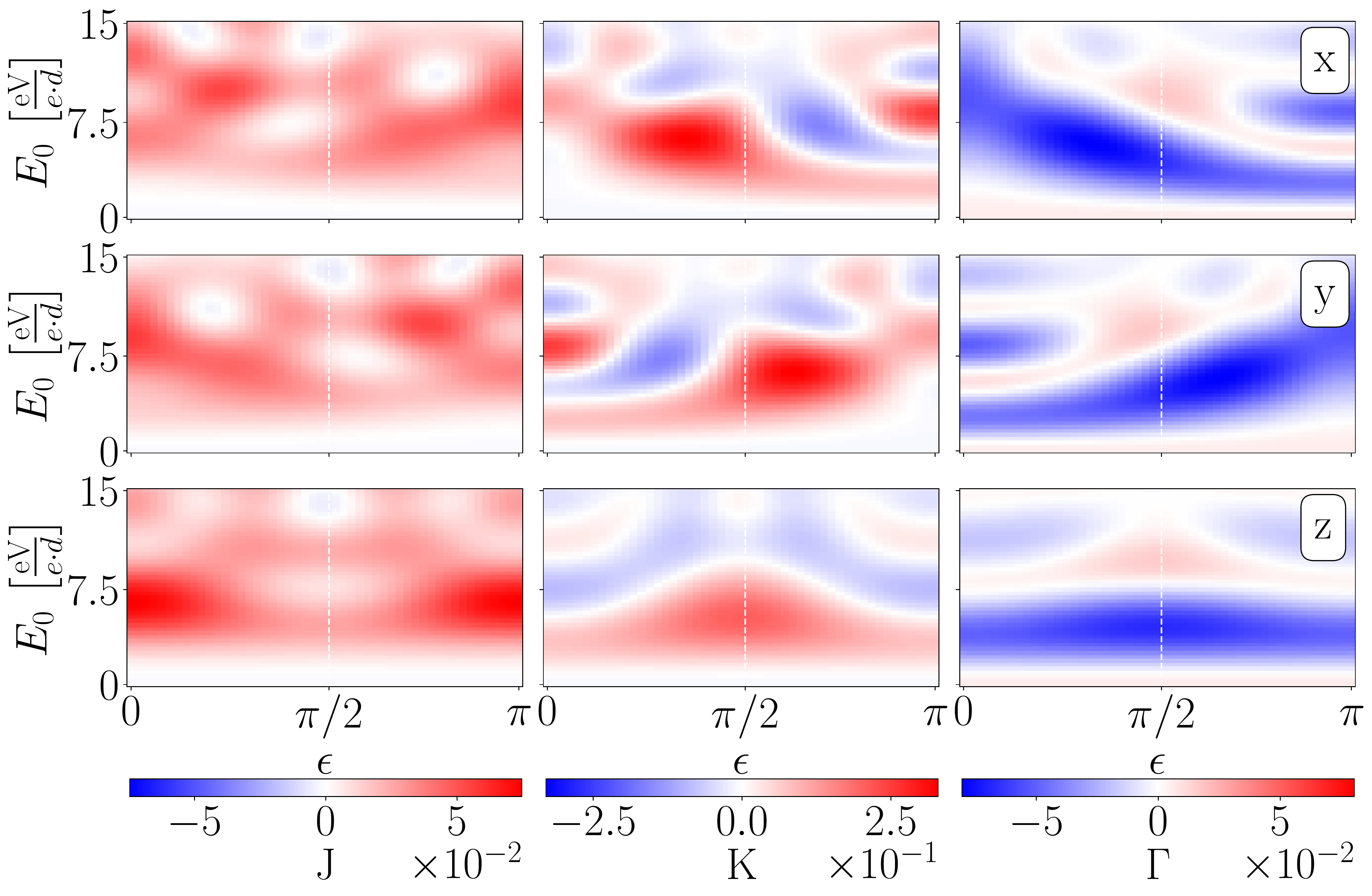}}
\subfigure[$D$-, $\mu$-, and h-interactions]{\label{fig:Fig4b}\includegraphics[width=\columnwidth]{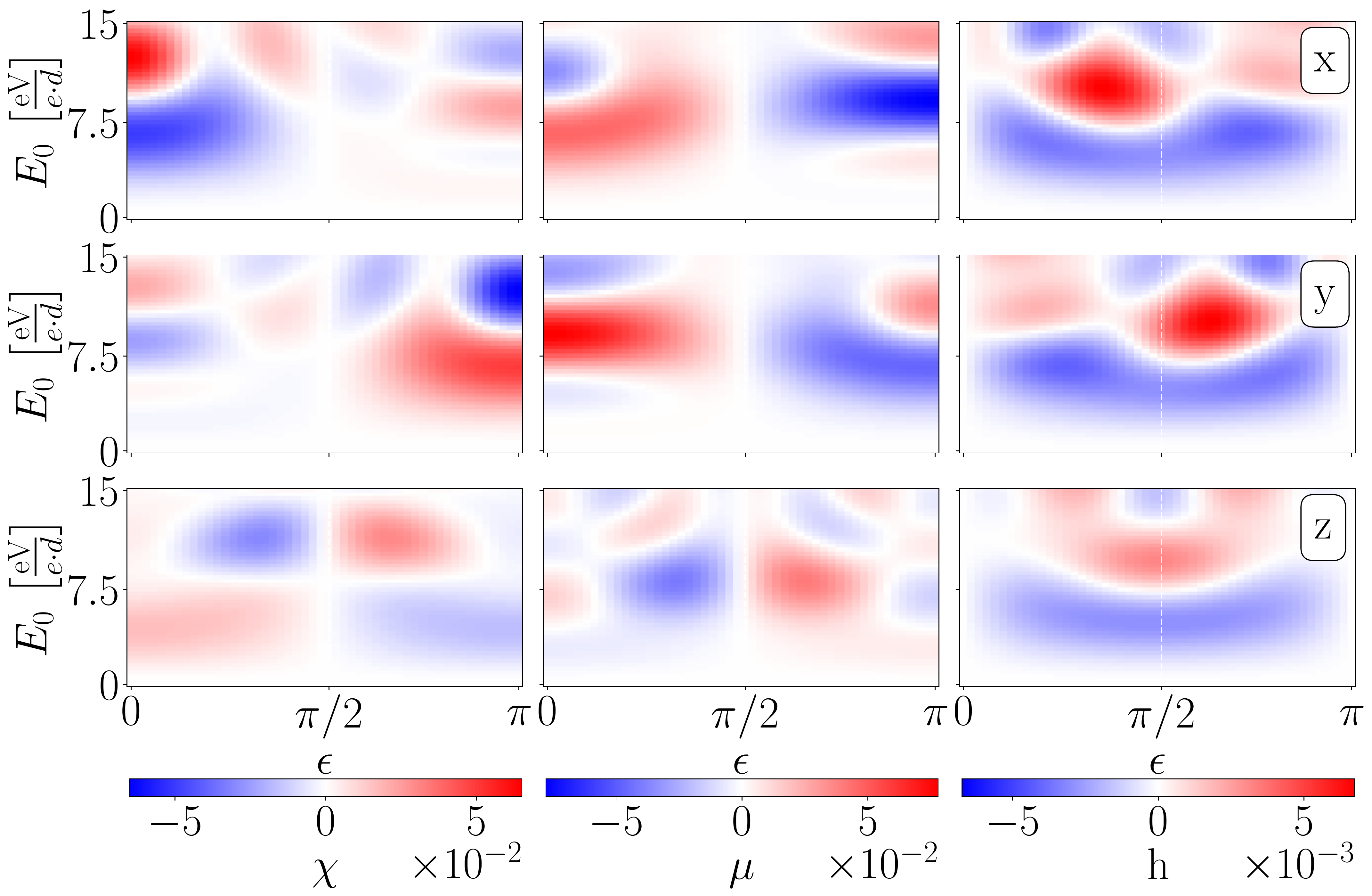}}
\caption{Interactions  in $x$-, $y$-, and $z$-direction in dependency of light amplitude $E_0$ and phase shift $\epsilon$ at $\omega=2.1\,$eV. Displayed are the results for frequency multiplicity N=1}\label{fig:Fig4}
\end{figure}

In this section we use the advantage of the Lissajous formalism (Sec.\ref{theo}), by continuously varying $\epsilon$ between the limiting cases introduced in Sec.\ref{LCA}. We analyze all interactions in dependency of $\epsilon$ and $E_0$, for $N=1$ and $\omega=12.0\,$eV in order to compare our results to the CPL results from~\cite{Kumar2022}. The results for all interactions are displayed in Fig.\ref{fig:Fig3}. 
 
As already discussed in Sec.\ref{theo} for CPL bond interactions are isotropic. Moving away from CPL the interactions become anisotropic immediately, i.e. if one desires to tune all bonds in the same manner one has to use CPL. For the J-, K-, and $\Gamma$ interaction [Fig.\ref{fig:Fig3a}] one observes a decrease in interaction strength for $E_0>10\,\mathrm{eV}/(e\cdot d)$ accompanied with the introduction of some sizable anisotropies moving from $\pi/2$ to $0$. For the $z$-bond $K$ and $J$-interactions we observe a change in sign for finite $E_0$. $\Gamma$ interactions are mainly suppressed for sizable $E_0$ for $\omega=12.0\,$eV. 

As already reported for CPL $D$ and $\mu$ vanish. However tuning $\epsilon\to 0$ induces finite values for both $D$ and $\mu$. We observe that $D$ and $\mu$ are anisotropic through the entire parameter range. Contributions become strong around the limit of LPL. We note that the $z$-term of $D$ and $\mu$ is antisymmetric around $\pi/2$, while other  $z$-interaction terms are symmetric around $\pi/2$. $x$- and $y$-interactions for $D$ and $\mu$ are also different from the other interactions. While for the latter $x$-interactions coincides with the $y$-interactions when changing $\epsilon\to -\epsilon$, for $D$ and $\mu$ there is a change in sign for the interactions. We attribute this behavior to the anisotropic nature of these interactions.  

Finally the magnetic field h, like for NLI interactions,  becomes anisotropic moving away from CPL. This can be interpreted as a change in direction of the induced magnetic field. While for CPL the direction is $\mathbf{n}$, for $\pi/4$ and $E_0\approx 20\,eV/(e\cdot d)$ the magnetic field points mainly in $x$-direction. Increasing the driving amplitude up to $E_0>40\,\mathrm{eV}/(e\cdot d)$, $x$ and $y$ contributions of the magnetic field vastly decrease and we obtain an induced magnetic field which points mainly out of plane.

For frequencies far above all resonances ($E_P, E_D,$ and $E_S$) like $\omega=12.0\,$eV we generally expect light to suppress interaction strength with increasing $E_0$. On the other hand frequencies between the resonances can induce a significant increase in interaction strength. This effect has been already reported for both CPL~\cite{PhysRevB.103.L100408,Kumar2022} as well as LPL~\cite{PhysRevB.104.214413,PhysRevB.105.085144,PhysRevResearch.4.L032036} and appears to be a promising route to obtain a KSL ground state. We therefore change the driving frequency to $\omega=2.1\,$eV which is between the resonances of $E_P$, $E_S/2$, and $\Delta$ (see App.~\ref{AD}) and evaluate all interactions. 

It has to be mentioned that driving between resonances comes with an increased risk of heating~\cite{PhysRevB.105.085144,PhysRevLett.121.107201}, and frequencies have to be chosen cautiously. A poor choice of frequencies might well give misleading results, since we are working in the ORA~\cite{PhysRevB.105.085144} which diverges around resonances. This is especially true for fourth-order terms which go with $1/\Delta^2$ i.e. frequencies close to the $\Delta$ resonance diverge even faster. (A detailed analysis of the resonances is given in the App.~\ref{AD}.) Results for $\omega=2.1\,$eV are shown in Fig.\ref{fig:Fig4a} and Fig.\ref{fig:Fig4b} for non-light-induced (NLI) interactions J, K, and $\Gamma$ and light-induced (LI) interactions  h,$D$, and $\mu$ respectively. 

The NLI interactions are significantly enhanced for finite $E_0$ throughout the whole range of $\epsilon$ considered. Furthermore we notice that the degree of enhancement strongly depends on $\epsilon$. For the Heisenberg interaction, the largest interactions can be found for LPL. Meanwhile for $K$- and $\Gamma$-interactions maxima can be found between LPL and CPL for $E_0\approx 7\,\mathrm{eV}/(e\cdot d)$ . As discussed earlier increasing Kitaev interactions compared to the other interactions is desirable and while this is possible via tuning $\epsilon$ it comes at the cost of loss of isotropy. Hence the model is not longer in the ideal Kitaev-Heisenberg picture of~\cite{PhysRevLett.102.017205,PhysRevLett.112.077204}, but could point more towards a dimerization of the ground state or a gapped spin liquid~\cite{KITAEV20062}. The fact that this is especially possible in between LPL and CPL emphasizes the importance of going beyond the limiting cases.

Looking at the LI interactions we observe significant increase in $D$- and $\mu$-contributions. For $\omega=12.0\,$eV both $D$ and $\mu$ were multiple orders smaller than the NLI interactions. Meanwhile for $\omega=2.1\,$eV, LPL, and $E_0>4\,\mathrm{eV}/(e\cdot d)$ their contributions increase noticeable, having the same magnitude as the NLI interactions. We observe that both $D_z$ and $\mu_z$ are antisymmetric around $\pi/2$, as it was the case for $\omega=12.0\,$eV. The induced magnetic field does not increase significantly compared to the results of $\omega=12.0\,$eV. Like for $\omega=12.0\,$eV we note that the maximal induced magnetic field in $x$- and $y$-direction is in between LPL and CPL at $\epsilon\approx \pi/3$.  
    
\subsection{$N>1$ Lissajou figures between resonances}\label{LCC}
\begin{figure}
\subfigure[J-, K-, and $\Gamma$-interactions]{\label{fig:Fig5a}\includegraphics[width=\columnwidth]{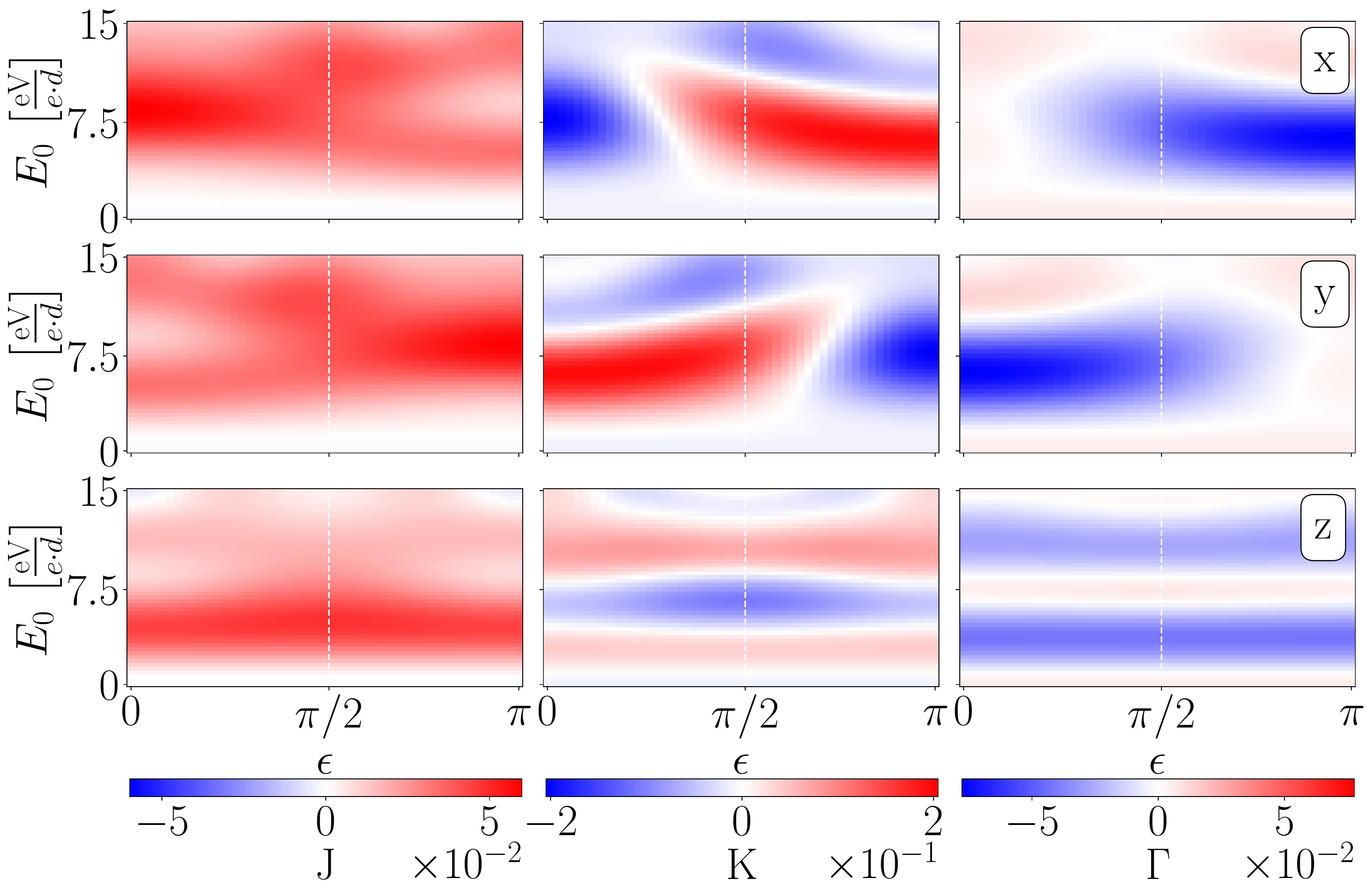}}
\subfigure[$D$-, $\mu$-, and h-interactions]{\label{fig:Fig5b}\includegraphics[width=\columnwidth]{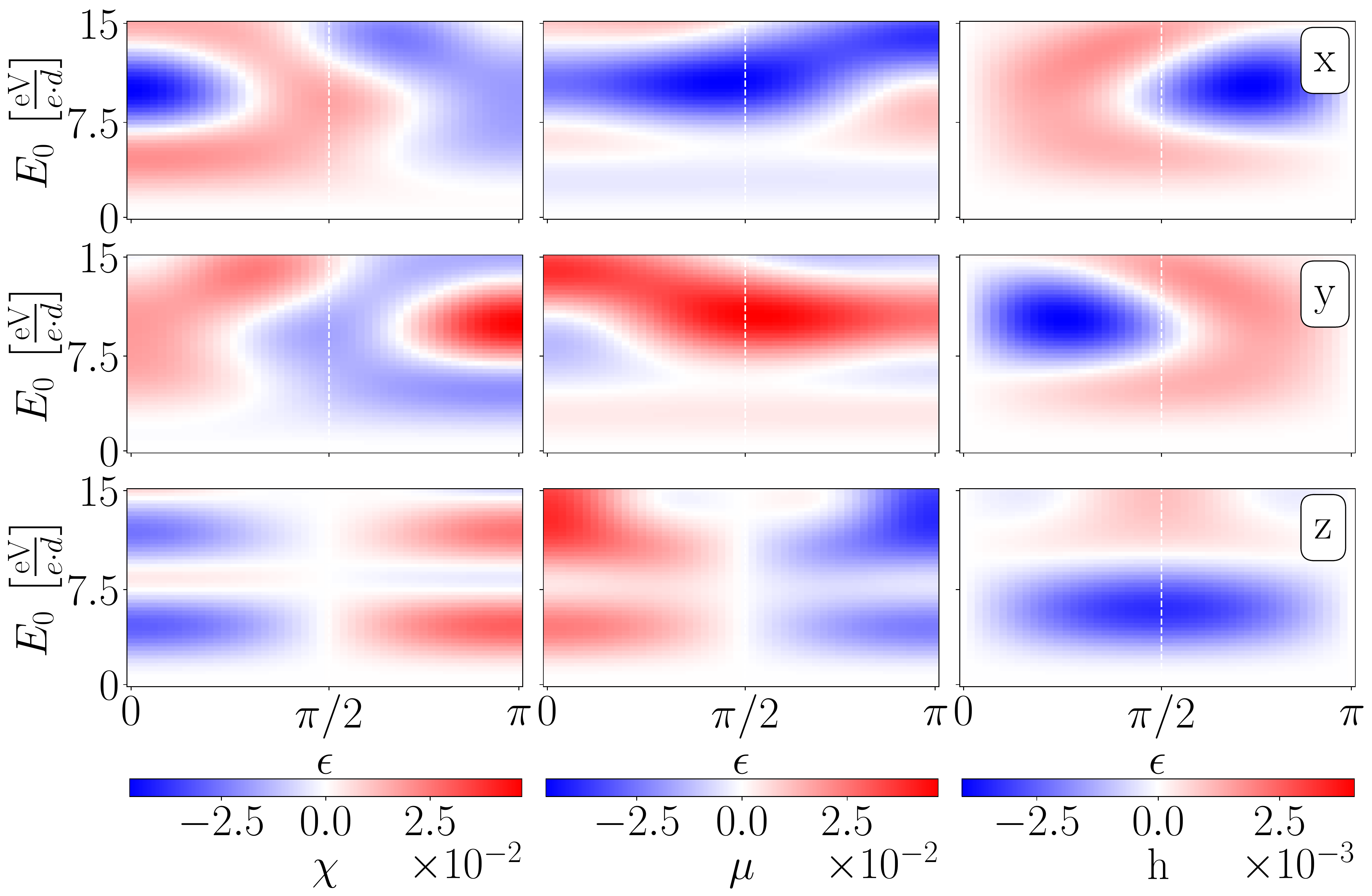}}
\caption{Interactions  in $x$-, $y$-, and $z$-direction in dependency of light amplitude $E_0$ and phase shift $\epsilon$ at $\omega=2.1\,$eV. Displayed are the results for frequency multiplicity N=2}
\label{fig:Fig5}
\end{figure} 

\begin{figure}
\subfigure[J-, K-, and $\Gamma$-interactions]{\label{fig:Fig6a}\includegraphics[width=\columnwidth]{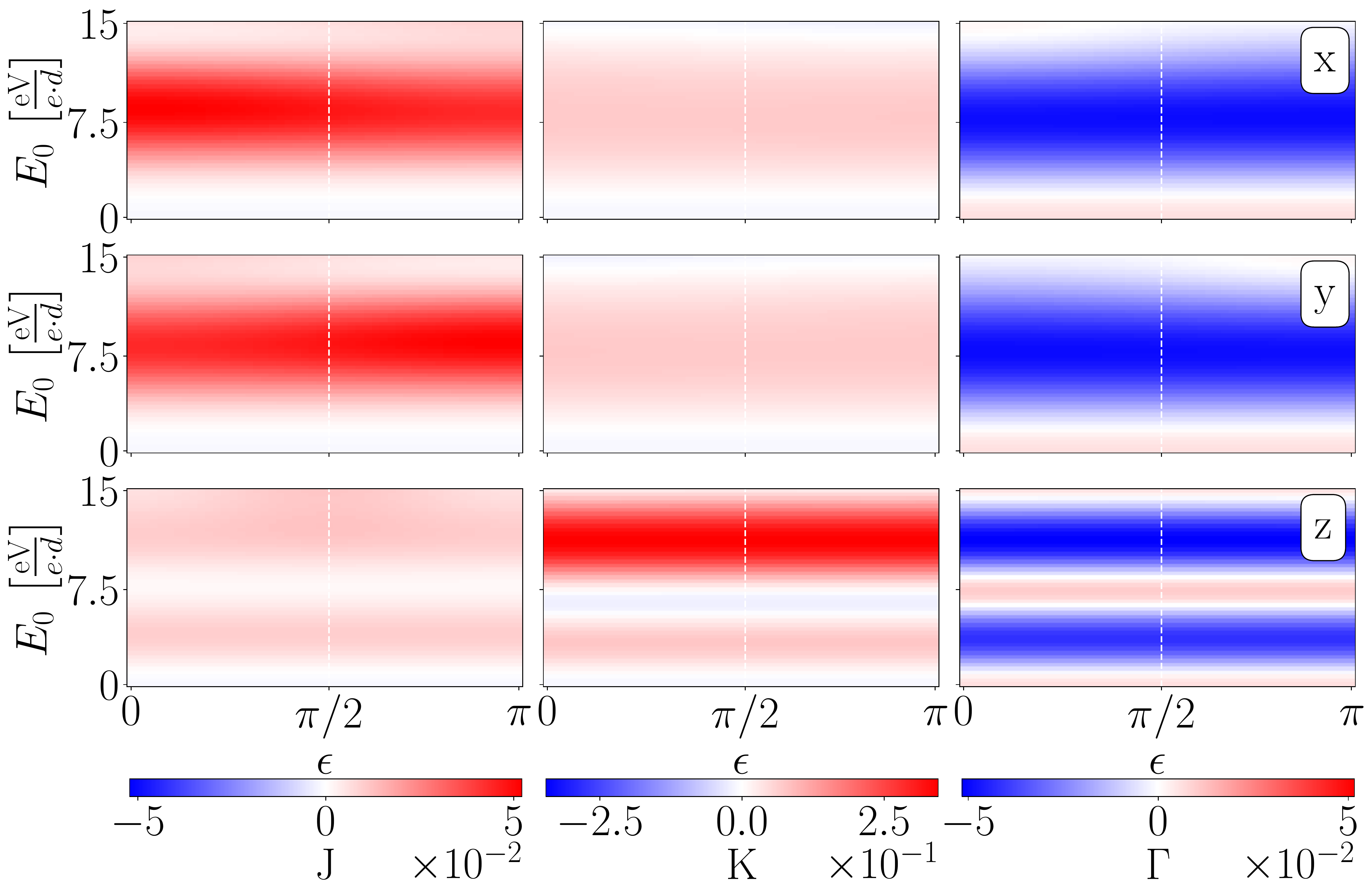}}
\subfigure[$D$-, $\mu$-, and h-interactions]{\label{fig:Fig6b}\includegraphics[width=\columnwidth]{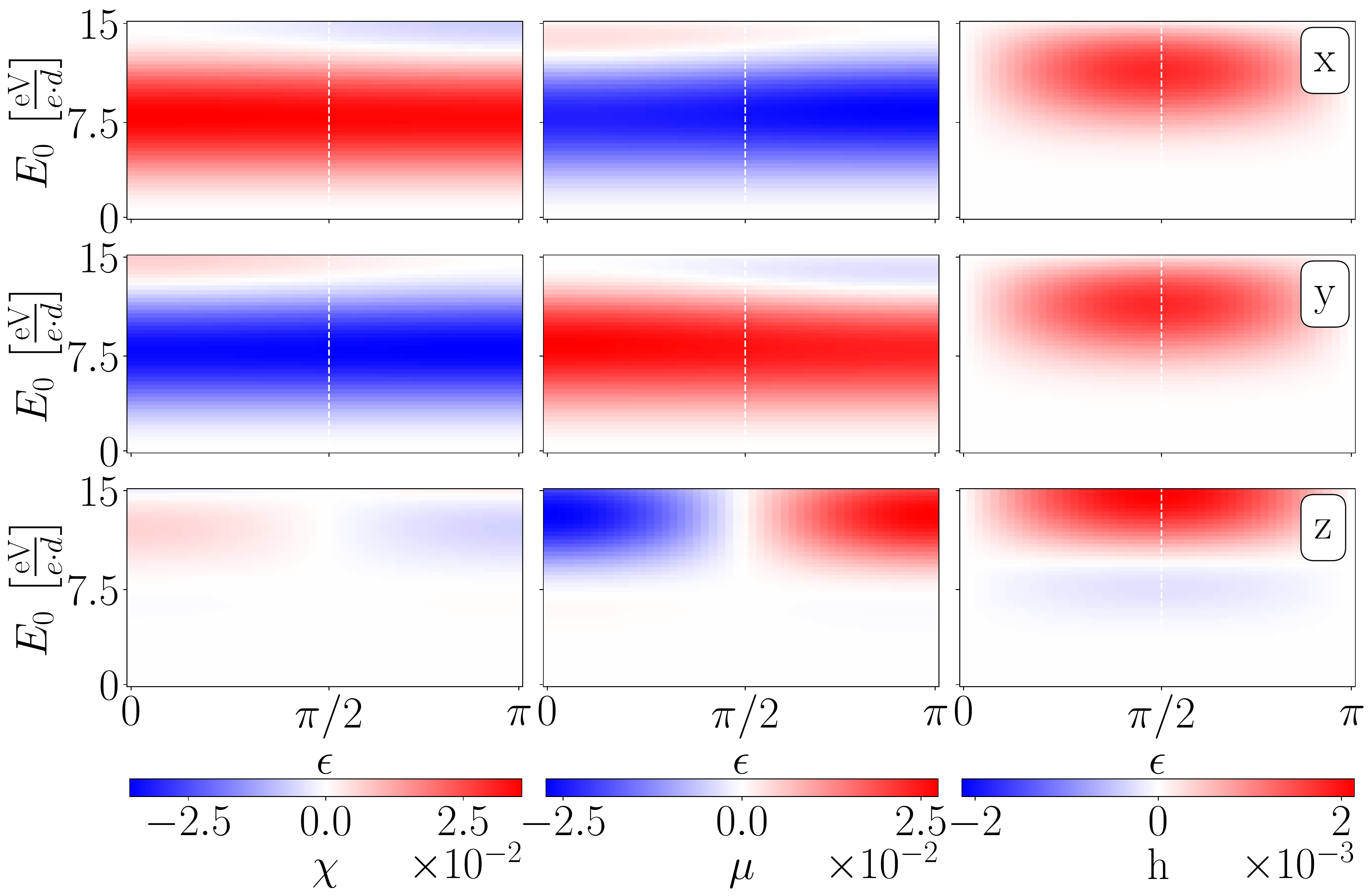}}
\caption{Interactions  in $x$-, $y$-, and $z$-direction in dependency of light amplitude $E_0$ and phase shift $\epsilon$ at $\omega=2.1\,$eV. Displayed are the results for frequency multiplicity N=5}
\label{fig:Fig6}
\end{figure} 

After analyzing influence of both $\epsilon$ and $\omega$ we want to discuss the influence of the frequency multiplicity $N$ in (\ref{eq:EQ15}) and (\ref{eq:EQ19}). As we have seen in Sec.\ref{LCB} and~\cite{PhysRevB.105.085144}, driving between resonances is the most promising pathway to increase Kitaev interactions. Therefore we set the driving frequency to $\omega=2.1\,$eV for the remainder of this section.

We start with $N=2$ Lissajous figures. The results for all interaction terms are displayed in Fig.\ref{fig:Fig5}. NLI interactions [Fig.\ref{fig:Fig5a}] show a distinct $\epsilon$- and $E_0$-dependency for $N=2$. However the magnitude of the interactions stays relatively unaffected. The most notable change is that for K and $\Gamma$ the maximal interaction strength arises for LPL. Furthermore, for $\pi/2$, not all bond interactions are isotropic anymore. While $x$- and $y$-interactions still obey the same $E_0$-dependence, $z$-interactions clearly differ from that. This is the case because $\epsilon=\pi/2$ and $N=2$ does not correspond to CPL but to a more complex Lissajous figure, with some anisotropy (see App.~\ref{AE}). The behavior of the $K_x$- and $K_y$-interaction, yields the possibility to switch signs of $x$- and $y$-interactions with changing $\epsilon$ from $0\to \pi$ at $E_0\approx 7.5\,eV/(e\cdot d)$. For the $\Gamma$-interactions this change suppresses $y$-interactions and enhances $x$-interactions. For both $K$- and $\Gamma$-interactions the $z$-bond is relatively unaffected by the changes in $\epsilon$. 

Contrary to CPL, LI interactions show non-zero contributions for $\epsilon=\pi/2$ and finite $E_0$. Last but not least the induced magnetic field $h$ has a comparable magnitude to the results of Fig.\ref{fig:Fig4b}, however $\epsilon$- and $E_0$-dependence vastly change. For $E_0>7.5\,\mathrm{eV}/(e\cdot d)$ in plane contributions are far more dominant than the $z$-interactions indicating an induced in-plane magnetic field. For smaller amplitudes $E_0<7.5\,\mathrm{eV}/(e\cdot d)$ we see a tendency for a magnetic field pointing out-of-plane.

In order to represent the behavior of the interactions for large frequency multiplicity $N$ we set $N=5$. The NLI interactions, displayed in Fig.\ref{fig:Fig6a}, almost completely decouple from the parameter $\epsilon$. This goes hand in hand with an almost isotropic behavior for the $x$- and $y$-bond interactions, with a distinct behavior for the $z$-bond. 
As is evident in (\ref{eq:EQ2}) an increase in $N$ reduces the effect of $\epsilon$, which is why NLI interactions are barely affected by $\epsilon$. High $N$ therefore offer the possibility of tuning one bond direction respective to the others, not only for $\epsilon=0$ but for $0<\epsilon<\pi$. For LI interactions [Fig.\ref{fig:Fig6b}], we observe a quite distinct behavior for certain interactions. $x$- and $y$-interactions of $D$ and $\mu$ are not $\epsilon$-dependent. Meanwhile the $z$-interaction for both vanishes at $\epsilon=\pi/2$ and has opposite signs for $\epsilon=0$ and $\epsilon=\pi$. One can therefore switch signs of the $z$-interaction while keeping the other interactions almost unchanged. Finally the induced magnetic field vanishes for $\epsilon=0$ and $\epsilon=\pi$ and therefore is clearly intertwined with $\epsilon$. Significant magnetic fields, just arise for relatively high amplitudes. This provides the possibility to turn the magnetic field on and off while keeping the other interactions intact. The magnetic field can therefore be controlled via the phase shift parameter $\epsilon$.      

\section{Summary and Outlook}\label{sum}
In this paper we derived an effective Floquet Kitaev-Heisenberg Hamiltonian up to fourth-order perturbation theory, capturing the effects of driving the system with arbitrary polarization. In addition to known tuning possibilities~\cite{PhysRevB.104.214413,PhysRevB.105.085144}, the relative frequency ($N$) and phase ($\epsilon$) of $x-$ and $y-$ polarization become decisive factors for interaction strengths. With our model we are able to continuously go from LPL to CPL and investigate the behavior for this transition. In addition, due to the inclusion of $N$, we are able to capture the behavior of the interaction parameters for arbitrary Lissajous figures.

We showed that while CPL induces a magnetic field $h$ due to TRS breaking, LPL breaks the inversion symmetry and induces the not yet reported terms $D$ and $\mu$, which might induce interesting physical properties. Furthermore we studied the interaction terms for different parameter settings, investigating the influence of $\epsilon$, $E_0$ and $N$ as well as $\omega$. We showed a that in order to significantly increase Kitaev interactions it is desirable to drive the system with frequencies between resonances. 
Moving away from CPL induces anisotropies but also increases Kitaev interaction which might result in agapped KSL ground state~\cite{PhysRevB.105.085144}. 
For $N>1$ we found that $N=2$ results in a quite distinct behavior to $N=1$ with $D$ and $\mu$ being a factor throughout the whole parameter range of $\epsilon$. Furthermore we found that for higher $N$ we have a clear tendency of $J$, $K$, and $\Gamma$ decoupling from $\epsilon$, while LI interactions still show a $\epsilon$-dependency to certain degree. With this a tuning of LI interactions, especially $h$, while keeping $J$, $K$ and $\Gamma$ unchanged seems possible.
Including third and fourth-order terms in perturbation theory for linear polarized light is crucial because the terms $D$ and $\mu$ only appear in third and higher order. 

Introducing arbitrary polarization into the Floquet Hamiltonian opened the doors for a multitude of yet undiscovered tuning possibilities via a plethora of parameters, making it both a very interesting and challenging topic for the future. From our studies we conclude that states like a gapped KSL with induced magnetic field could be in the range of possibilities, tuning with $N\gg 1$ light. For the future it would be interesting to analyze the ground state arising from the interactions for both isotropic and anisotropic cases.


\bibliography{References}
\clearpage
\appendix 
\section{Second order effective Hamiltonian for arbitrary polarization}\label{AA}
\begin{figure}
\includegraphics[width=\textwidth]{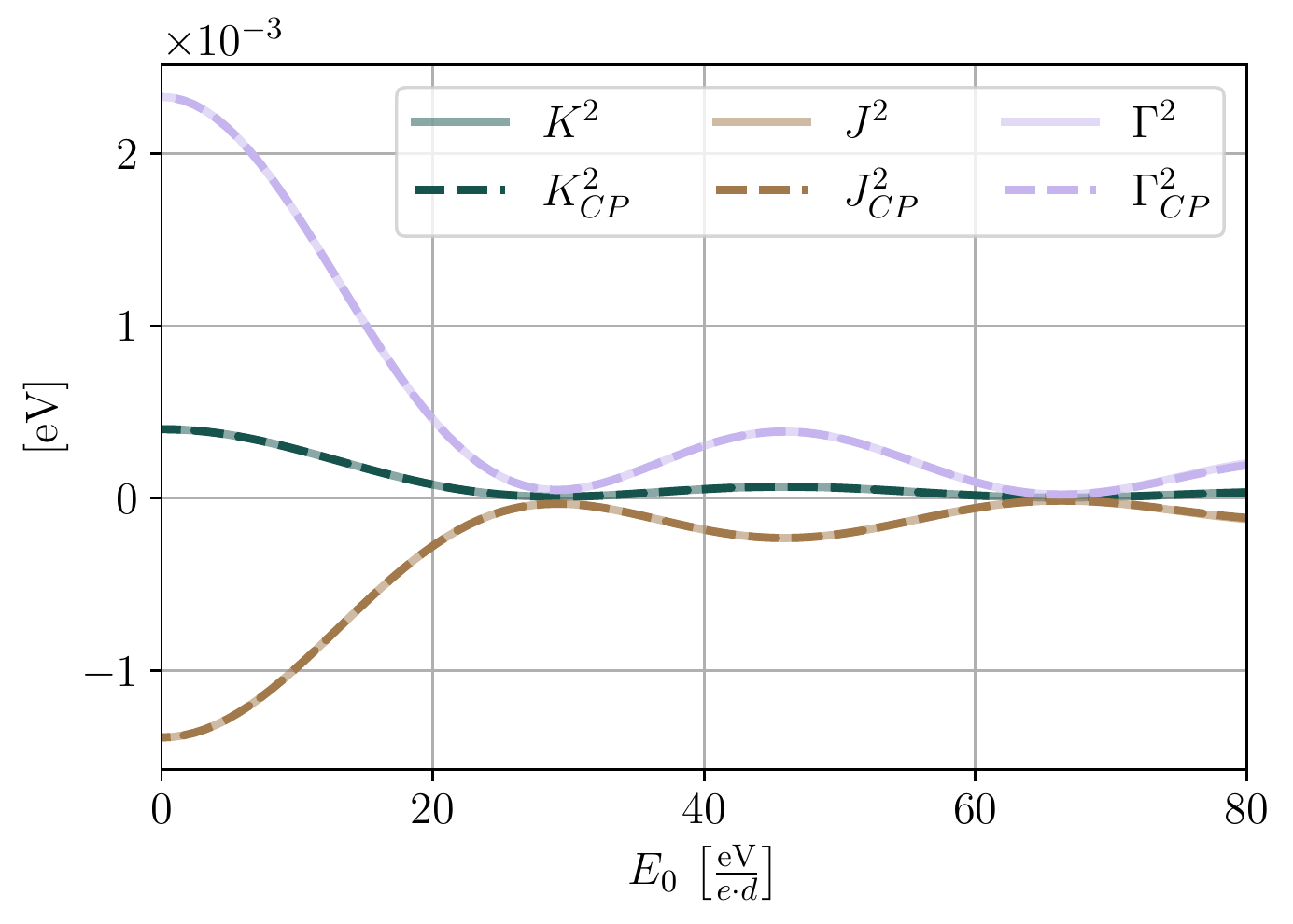}
\caption{Comparison of the model for CPL (dashed lines) with the second order results for arbitrary polarization (\ref{eq:A1})-(\ref{eq:A3}) in the limit of $N=1$ and $\epsilon=\pi/2$ (solid lines) at $\omega=12.0\,$eV}\label{supp:Fig1}
\end{figure}
As explained in Sec.\ref{theoB}, the second order Kitaev-Heisenberg model under the influence of arbitrary polarized light can be obtained by replacing the Bessel functions in~\cite{PhysRevB.105.085144} with $\mathcal{B_l}$, derived in Sec.\ref{theoA}. In (6a)-(6c) of~\cite{Kumar2022} we set $\mathcal{J}_l\mathcal{J}_l\rightarrow\mathcal{B}^{\ast}_l\mathcal{B}_l=|\mathcal{B}_l|^2$. $\mathcal{B}^{\ast}_l$ here describes the hopping back to the initial site with an emission of $l$ photons. Therefore the interactions $J, K,$ and $\Gamma$ become
\begin{align}
J^2=&\sum_{l=-\infty}^{\infty}|\mathcal{B}_l(\vartheta,\mathbf{A})|^2\frac{4}{27}\bigg(\frac{6t_1(t_1+2t_3)}{U-3J_H-l\omega}\notag\\&+\frac{2(t_1-t_3)^2}{U-J_H-l\omega}+\frac{(2t_1+t_3)^2}{U+2J_H-l\omega}\bigg)\label{eq:A1}\;,\\
K^2=&\sum_{l=-\infty}^{\infty}|\mathcal{B}_l(\vartheta,\mathbf{A})|^2\frac{4}{9}(t_1-t_3)^2-3t_2^2\notag\\&\times\left(\frac{1}{U-3J_H-l\omega}-\frac{1}{U-J_H-l\omega}\right)\;,\label{eq:A2}\\
\Gamma^2 =&\sum_{l=-\infty}^{\infty}|\mathcal{B}_l(\vartheta,\mathbf{A})|^2\frac{8}{9}t_2(t_1-t_3)\notag\\&\times\left(\frac{1}{U-3J_H-l\omega}-\frac{1}{U-J_H-l\omega}\right)\;,\label{eq:A3}
\end{align}
with $\mathcal{B}_l(\vartheta,\mathbf{A})$ from (\ref{eq:EQ3}). The term $|\mathcal{B}_l(\vartheta,\mathbf{A})|^2$ can be written as
\begin{widetext}
\begin{align}
|\mathcal{B}_l(\vartheta,\mathbf{A})|^2=\left[\sum_n\mathcal{J}_{l-Nn}\left(\frac{E_0}{\omega}\cos(\vartheta)\right)\mathcal{J}_{n}\left(\frac{E_0}{N\omega}\sin(\vartheta)\right)\cos(\epsilon n)\right]^2+\left[\sum_n\mathcal{J}_{l-Nn}\left(\frac{E_0}{\omega}\cos(\vartheta)\right)\mathcal{J}_{n}\left(\frac{E_0}{N\omega}\sin(\vartheta)\right)\sin(\epsilon n)\right]^2.
\label{eq:A4}
\end{align} 
\end{widetext}
For $\epsilon=\pi/2$ and $N=1$ equations (\ref{eq:A1})-(\ref{eq:A3}) have to coincide with (6a)-(6c) of~\cite{Kumar2022}. In Fig.\ref{supp:Fig2} we display the results for CPL and arbitrary polarization for $N=1$ and $\epsilon=\pi/2$ for $\omega=12.0\,$eV. The results of the CPL model~\cite{Kumar2022,PhysRevB.103.L100408} are displayed with dashed lines, while the results of (\ref{eq:A1})-(\ref{eq:A3}) are shown in solid lines. As expected, we observe a perfect agreement between both models. 

\section{$\epsilon=\pi/2$ - Circular Polarized Light}\label{AB}
 Most studies up until now have focused on the influence of CPL on $\alpha$-RuCL$_3$, due to the induced inverse Faraday effect as well as the uniform influence of the light on all bonds. In this section we will therefore study circular polarized light as a special case of third (\ref{eq:EQ10})-(\ref{eq:EQ14}) and fourth order results (\ref{eq:EQ16})-(\ref{eq:EQ18}) for arbitrary polarization. In order to describe circular polarized light we have to set $N=1$ and $\epsilon=\pi/2$ in (\ref{eq:EQ1.5}). With this, expressions for third and fourth order terms become significantly easier to handle.  

Starting with third order terms [(\ref{eq:EQ10})-(\ref{eq:EQ14})] we obtain
\begin{widetext}
\begin{align}
K^3=&\sum_{m,l}\mathfrak{J}^3_{m,l}(A_0)\frac{t_{pd}^2}{\Delta+m\omega}\bigg[\frac{12}{9}\cos\left[(m-n)\frac{\pi}{4}\right]\left(\frac{t_2}{E_D+(l+m)\omega}-\frac{t_2}{E_P+(l+m)\omega}\right)
\notag\\
&+\sin\left[(m-n)\frac{\pi}{4}\right]\frac{8}{27}\left(6\frac{t_1}{E_P+(l+m)\omega}+\frac{t_1-t_3}{E_D+(l+m)\omega}+\frac{2t_1+t_3}{E_S+(l+m)\omega}\right)\bigg]\label{eq:A5}\\
\Gamma^3=&\sum_{m,l}\mathfrak{J}^3_{m,l}(A_0)\frac{t_{pd}^2}{\Delta+m\omega}\frac{4}{9}\cos\left[(m-n)\frac{\pi}{4}\right]\left(\frac{t_1-t_3}{E_P+(l+m)\omega}-\frac{t_1-t_3}{E_D+(l+m)\omega}\right)\label{eq:A6}\\
h^3=&\sum_{m,l}\mathfrak{J}^3_{m,l}(A_0)\frac{-t_{pd}^2}{\Delta+m\omega}\frac{2}{9}\sin\left[(m-n)\frac{\pi}{4}\right]\left(\frac{t_1-t_3}{E_D-(l+m)\omega}+\frac{t_1-t_3}{E_P+(l+m)\omega}\right),\label{eq:A7}
\end{align}
\end{widetext}
with $\mathfrak{J}^3_{m,l}(A_0)=\mathcal{J}_{m+l}\left(A_0\right)\mathcal{J}_{l}\left(\tilde{A}_0\right)\mathcal{J}_{m}\left(\tilde{A}_0\right)$, where $\tilde{A}_0=A_0/\sqrt{2}$.
We observe that both $\mu$ and $D$ vanish. Meanwhile $h^3$, which breaks time reversal symmetry, prevails. For this reason the terms $\chi$ and $g$ have not been reported in the studies on circular polarized light~\cite{PhysRevResearch.4.L032036,PhysRevResearch.4.L032032,Kumar2022}. For the absence of light, i.e. $A_0=0$, we have only contributions for $m=l=0$, which means that $\sin[(m-n)\pi/4]=0$ in (\ref{eq:A5})-(\ref{eq:A7}) while $\cos[(m-n)\pi/4]=1$. Therefore the magnetic field vanishes in absence of light and the contributions of (\ref{eq:A5}) and (\ref{eq:A6}) reproduce exactly the result for non-Floquet perturbation theory~\cite{PhysRevLett.112.077204}. 

Analogue to the third order results (\ref{eq:A5})-(\ref{eq:A7}) the special case of CPL for fourth order terms [(\ref{eq:EQ16})-(\ref{eq:EQ18})] yields
\begin{widetext}
\begin{align}
J^4=&\sum_{k,l,m}\frac{t_{pd}^4}{(\Delta+(l+m+k)\omega)(\Delta+m\omega)}\mathfrak{J}_{m,k,l}^4(A_0)\frac{2}{27}\left(\frac{2}{E_S+(m+k)\omega}+\frac{3}{E_P+(m+k)\omega}+\frac{1}{E_D+(m+k)\omega}\right)\notag\\&\times\left[\cos\left((l-k)\frac{\pi}{2}\right]-\cos\left[(m-l)\frac{\pi}{2}\right]\right)\label{eq:A9}\\
K^4=&\sum_{k,l,m}\frac{t_{pd}^4}{(\Delta+(l+m+k)\omega)(\Delta+m\omega)}\mathfrak{J}_{m,k,l}^4(A_0)
\bigg[\frac{2}{3}\left(\frac{1}{E_P+(m+k)\omega}-\frac{1}{E_D+(m+k)\omega}\right)\cos\left[(m-l)\frac{\pi}{2}\right]\notag\\
&-\frac{2}{27}\left(\frac{2}{E_S+(m+k)\omega}+\frac{3}{E_P+(m+k)\omega}+\frac{4}{E_D+(m+k)\omega}\right)\left(\cos\left[(l-k)\frac{\pi}{2}\right]-\cos\left[(m-l)\frac{\pi}{2}\right]\right)\bigg],\label{eq:A10}
\end{align}
\end{widetext}
with $\mathfrak{J}_{m,k,l}^4(A_0)=\mathcal{J}_k\left(\tilde{A}_0\right)\mathcal{J}_l\left(\tilde{A}_0\right)\mathcal{J}_m\left(\tilde{A}_0\right)\mathcal{J}_{m+k+l}\left(\tilde{A}_0\right)$.
Like for the third order terms $\mu$ vanishes for CP light, while corrections for Kitaev and Heisenberg interactions prevail. These correction terms in combination with the left out third order processes discussed in Sec.\ref{theoC} explain the discrepancies between numerical and analytical results in~\cite{Kumar2022}. For $A_0=0$ the Heisenberg interactions vanish while Kitaev interactions are finite, coinciding with the results of~\cite{PhysRevLett.112.077204}.  

\section{Matrix elements for interactions}\label{AC}
In order to project the effective spin-orbital Hamiltonian into the $j=1/2$ basis we have to calculate the matrix elements, determining all interaction parameters,
\begin{align}
J=&2\,\mathrm{Re}\left(\bra{\frac{1}{2},-\frac{1}{2}}H_{\mathrm{eff}}\ket{-\frac{1}{2},\frac{1}{2}}\right)\label{eq:C1}\\
\chi&=2\mathrm{Im}\left(\bra{\frac{1}{2},-\frac{1}{2}}H_{\mathrm{eff}}\ket{-\frac{1}{2},\frac{1}{2}}\right)\label{eq:C2}\\
h=&\frac{1}{2}\left(\bra{\frac{1}{2},\frac{1}{2}}H_{\mathrm{eff}}\ket{\frac{1}{2},\frac{1}{2}}-\bra{-\frac{1}{2},-\frac{1}{2}}H_{\mathrm{eff}}\ket{-\frac{1}{2},-\frac{1}{2}}\right)\label{eq:C3}\\
K=&\bra{\frac{1}{2},\frac{1}{2}}H_{\mathrm{eff}}\ket{\frac{1}{2},\frac{1}{2}}+\bra{-\frac{1}{2},-\frac{1}{2}}H_{\mathrm{eff}}\ket{-\frac{1}{2},-\frac{1}{2}}\notag\label{eq:C4}\\
&-2\bra{-\frac{1}{2},\frac{1}{2}}H_{\mathrm{eff}}\ket{-\frac{1}{2},\frac{1}{2}}-J\\
\Gamma&=-2\,\mathrm{Im}\left(\bra{\frac{1}{2},\frac{1}{2}}H_{\mathrm{eff}}\ket{-\frac{1}{2},-\frac{1}{2}}\right)\label{eq:C5}\\
\mu=&2\,\mathrm{Re}\left(\bra{\frac{1}{2},\frac{1}{2}}H_{\mathrm{eff}}\ket{-\frac{1}{2},-\frac{1}{2}}\right)\label{eq:C6}.
\end{align}
The results are similar to~\cite{Kumar2022}, with the distinction that calculating the $\bra{\frac{1}{2},-\frac{1}{2}}H_{\mathrm{eff}}\ket{-\frac{1}{2},\frac{1}{2}}$ and $\bra{\frac{1}{2},\frac{1}{2}}H_{\mathrm{eff}}\ket{-\frac{1}{2},-\frac{1}{2}}$ elements with all hopping processes results in real and imaginary contributions. Here the element (\ref{eq:C6}) yields the $\Gamma$ interactions and (\ref{eq:C1}) yields the Heisenberg interaction, just like in~\cite{Kumar2022}. The contributions (\ref{eq:C2}) and (\ref{eq:C6}) can neither be categorized as $\Gamma$ or $J$ interactions. In order to find a physical interpretation of these contributions we mapped them back into the $x,y,z$ basis yielding
\begin{align}
\left(\mathbf{S}_i\times\mathbf{S}_j\right)_z=&S_i^xS_j^y-S_i^yS_j^x=\frac{1}{2i}\left(S^-_iS^+_j-S^+_iS^-_j\right)\\
&S^x_iS^x_j-S^y_iS^y_j=\frac{1}{2}\left(S_i^+S_j^++S^-_iS^-_j\right)
\end{align}
which are the $D$ interactions, breaking IS, and $\mu$ interactions, inducing further anisotropies.
\section{Resonance frequencies}\label{AD}
\begin{figure}
\includegraphics[width=0.8\textwidth]{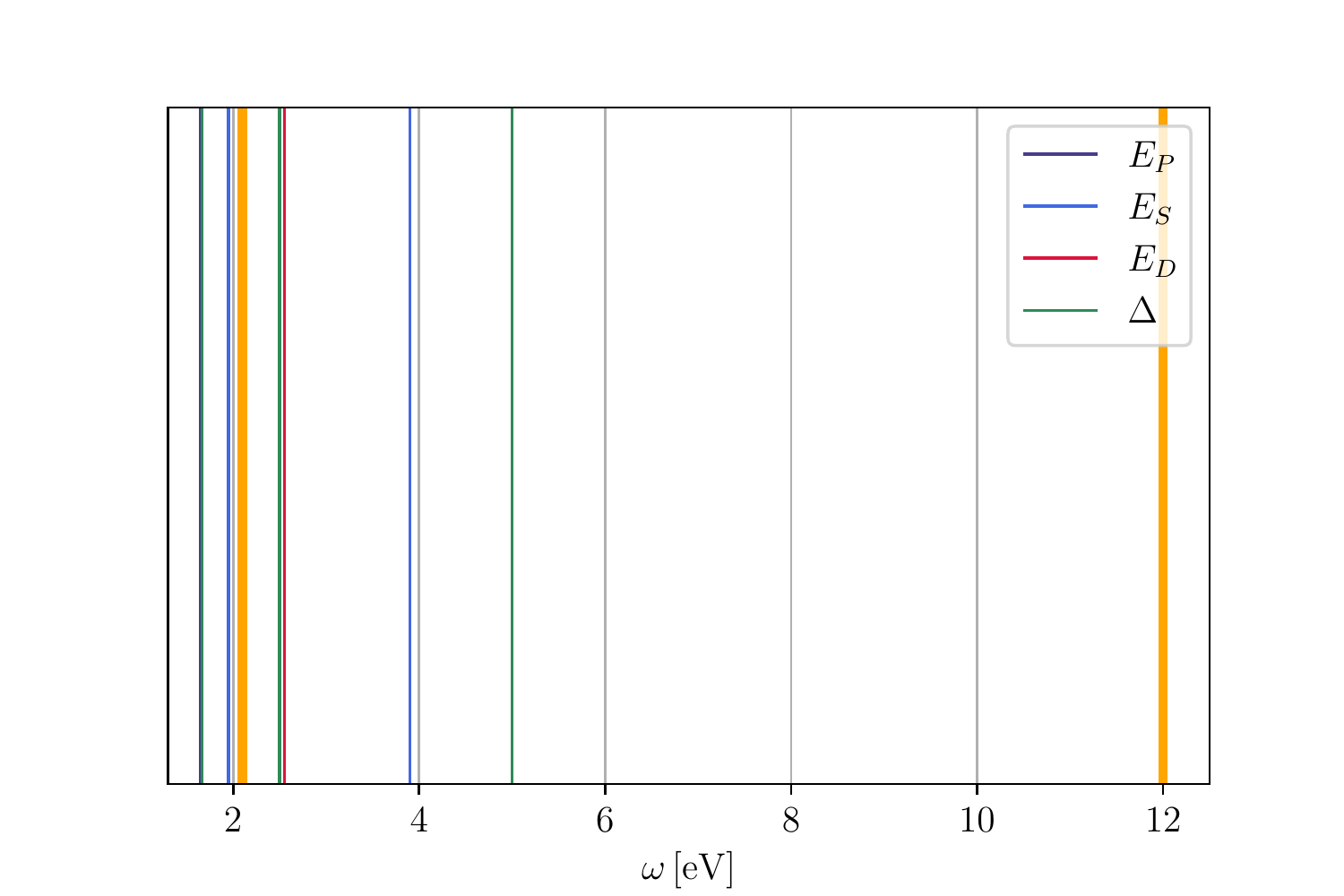}
\caption{Resonances of the effective Floquet Kitaev Heisenberg model. The resonces attributed to $E_S$, $E_P$, $E_D$, and $\Delta$ are displayed in blue, puprle , red, and green respectively. The driving frequencies used in the main text are visualized with orange lines.}\label{supp:Fig2}
\end{figure}
The resonance frequencies of the system are integer multiples of the excitation energies. In addition to  $E_S, E_D$, and $E_P$~\cite{PhysRevB.105.085144} we also have to consider $\Delta$, since we are including the $p$-ligands explicitly in our calculation. The resonances for the considered parameter setting of $U=3.0\,$eV and $J_H=0.45\,$eV are displayed in Fig.\ref{supp:Fig2} as black lines. The driving frequencies chosen in the main article are visualized as orange lines. The driving frequency $\omega=12.0\,$eV is clearly above all resonances. Meanwhile $\omega=2.1\,$eV lies in between the $E_S/2$, $E_D$, and $\Delta/2$ resonances. $E_D$ and $\Delta/2$ almost coincide, which leads to more pronounced heating effects~\cite{PhysRevB.105.085144,PhysRevLett.121.107201} close to this quasi-double resonance. We chose the driving frequency not exactly between $E_S/2$ and the double resonance but a little bit closer to the $E_S/2$ resonance, in order to work with less pronounced heating effects. 
\section{Lissajous figures}\label{AE}
\begin{figure}
\includegraphics[width=0.8\textwidth]{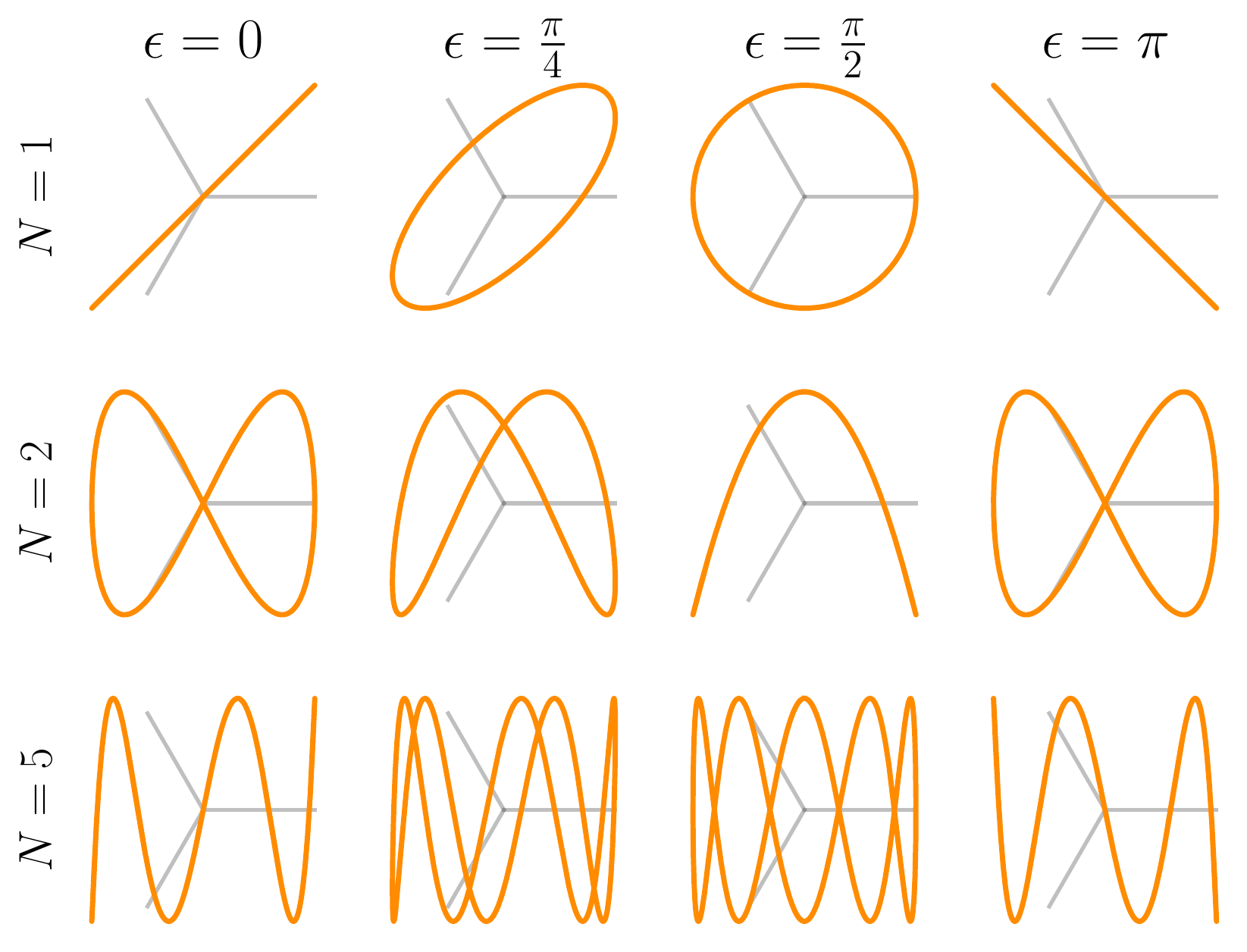}
\caption{Snapshots of Lissajous figures discussed in the main text. Displayed are Lissajous figures for $\epsilon=0,\,\pi/4,\,\pi/2,$ and $\pi$ for $N=1,\,2,$ and $5$ respectively in orange. The bond directions $x$, $y$, and $z$ are showcased in grey.}
\label{Supp:Fig3}
\end{figure}
Fig.\ref{Supp:Fig3} displays snapshots from a few Lissajous figures discussed in the main text. It becomes evident that the only Lissajous figure with a uniform influence  on all bond directions (grey lines in Fig.\ref{Supp:Fig3}) is $N=1$ and $\epsilon=\pi/2$. The reason for the opposing behavior of $x$ and $y$ interactions for $N=1$ can be understood looking at the Lissajous figure for $\epsilon=\pi$ and $\epsilon=0$. $\epsilon=\pi$ is LPL rotated by $\pi/2$ compared to $\epsilon=0$, this makes the influence of $\epsilon=0$ on the $x$-bond the same as $\epsilon=\pi$ on the $y$-bond, which results in the opposing behavior reported in the main text. 

\end{document}